\documentclass[useAMS,usenatbib]{mn2e}

\usepackage{epsf,rotating}
\usepackage{amsmath}
\usepackage{subfig}

\newcommand{\kmsmpc}{\kms\;{\rm Mpc}^{-1}}

\newcommand{\HI}{\ion{H}{i}}

\newcommand{\hkpc}{h^{-1}{\rm kpc}}
\newcommand{\hmpc}{h^{-1}{\rm Mpc}}

\newcommand{\kms}{\;{\rm km}\,{\rm s}^{-1}}

\newcommand{\gad}{{\sc Gadget-3}}

\newcommand{\mufasa}{{\sc Mufasa}}

\newcommand{\ion}[2]{\hbox{#1\,{\sc #2}}}

\title[Star formation, gas, \& metals in {\sc Mufasa}]{{\mufasa}:
Galaxy star formation, gas, and metal properties across cosmic time}

\author[Dav\'e et al.]{
\parbox[t]{\textwidth}{\vspace{-1cm}
Romeel Dav\'e$^{1,2,3}$, Mika H. Rafieferantsoa$^{1,5}$, Robert J. Thompson$^{4,1}$, Philip F. Hopkins$^{6}$}
\\
\\$^1$ University of the Western Cape, Bellville, Cape Town 7535, South Africa
\\$^2$ South African Astronomical Observatories, Observatory, Cape Town 7925, South Africa
\\$^3$ African Institute for Mathematical Sciences, Muizenberg, Cape Town 7945, South Africa
\\$^4$ National Center for Supercomputing Applications, Champaign-Urbana, IL 61801
\\$^5$ Max-Planck-Instit\"ut f\"ur Astrophysik, Garching, Germany
\\$^6$ California Institute of Technology, Pasadena, CA 91125
}

\begin{document}

\maketitle

 \begin{abstract}
We examine galaxy star formation rates (SFRs), metallicities, and
gas contents predicted by the \mufasa\ cosmological hydrodynamic
simulations, which employ meshless hydrodynamics and novel feedback
prescriptions that yield a good match to observed galaxy stellar
mass assembly.  We combine $50, 25,$ and $12.5\hmpc$ boxes with a
quarter billion particles each to show that \mufasa\ broadly
reproduces a wide range of relevant observations, including SFR and
specific SFR functions, the mass-metallicity relation, \ion{H}{i}
and H$_2$ fractions, \ion{H}{i} (21~cm) and CO luminosity functions,
and cosmic gas density evolution.  There are mild but significant
discrepancies, such as too many high-SFR galaxies, overly metal-rich
and \ion{H}{i}-poor galaxies at $M_*\ga 10^{10} M_\odot$, and sSFRs
that are too low at $z\sim 1-2$.  The \ion{H}{i} mass function
increases by $\times 2$ out to $z\sim1$ then steepens to higher
redshifts, while the CO luminosity function computed using the
Narayanan et al. conversion factor shows a rapid increase of
CO-bright galaxies out to $z\sim 2$ in accord with data.  $\Omega_{HI}$
and $\Omega_{H2}$ both scale roughly as $\propto (1+z)^{0.7}$ out
to $z\sim 3$, comparable to the rise in \ion{H}{i} and H$_2$
fractions.  \mufasa\ galaxies with high SFR at a given $M_*$ have
lower metallicities and higher \ion{H}{i} and H$_2$ fractions,
following observed trends; we make quantitative predictions for how
fluctuations in the baryon cycle drive correlated scatter around
galaxy scaling relations.  Most of these trends are well converged
with numerical resolution.  These successes highlight \mufasa\ as
a viable platform to study many facets of cosmological galaxy
evolution.
\end{abstract}

\begin{keywords}
galaxies: formation, galaxies: evolution, galaxies: star formation, galaxies: abundances, galaxies: ISM, methods: numerical
\end{keywords}

\section{Introduction}

Observations of galaxy properties from today back to the early
universe are improving at a remarkable pace, thanks to advancing
multi-wavelength photometric and spectroscopic galaxy surveys.
Progress has been particularly impressive in the near-infrared and
longer wavelengths, which provides more robust constraints on stellar
and metal content at high redshifts and gas content across all
redshifts.  Models for galaxy formation thus find it increasingly
challenging to be able to reproduce such observations within a
physically-motivated concordance cosmology framework.

Recent cosmological hydrodynamic simulations have been impressively
successful at broadly reproducing key galaxy demographic observables
over cosmic time~\citep[see][and references therein]{Somerville-15}.
A primary benchmark used to test galaxy formation models is the
observed galaxy stellar mass function (GSMF).  Many modern simulations
can now match this to within a factor of several over the majority
of cosmic time and
mass~\citep{Dave-13,Genel-14,Crain-15,Khandai-15,Dave-16,Kaviraj-16}, which
is typically within the range of current systematic uncertainties
in the data.  To do so, all cosmological-scale simulations incorporate
heuristic models for feedback processes associated with star formation
that suppress galaxy formation at the low-mass end, combined with
feedback often associated with active galactic nuclei (AGN) that
suppresses massive galaxy growth.  However, the precise physical
mechanisms invoked for feedback can vary substantially amongst
simulations, despite their predicted GSMFs being similar.  To further
test and discriminate between models, and thereby constrain the
physical mechanisms giving rise to feedback, it is thus important
to move beyond the GSMF and consider other aspects of galaxy
demographics.

Advancing multi-wavelength observations have made impressive progress
at characterising the gas and metal content of galaxies across
cosmic time.  Metallicity measures at higher redshifts have been
aided by new near-IR spectroscopic capabilities that have enabled
the same optical emission line measures used at low redshifts to
be applied to $z\sim 2-3$ galaxies~\citep{Steidel-14,Sanders-15}.
Molecular gas contents have now been measured out to similar redshifts
thanks to deep millimetre-wave data that can detect redshifted
carbon monoxide (CO) emission lines~\citep{Geach-11,Tacconi-13}.
Direct measures of atomic gas (\ion{H}{i}) remain confined to low
redshifts ($z\la 0.5$) as of yet owing to the sensitivity of current
instruments \citep{Delhaize-13,Fernandez-16}, but the Square Kilometre
Array (SKA) and its prescursors such as MeerKAT aim to probe
\ion{H}{i} out to $z\sim 1$ and beyond~\citep[e.g.][]{Holwerda-12}.
These observations provide a direct glimpse into the gaseous fuel
for star formation, as well as products of massive star formation
as traced by chemical enrichment, hence they can more directly probe the
baryon cycle of gaseous inflows and outflows that are viewed as
being the central driver of cosmological galaxy evolution.

Cosmological galaxy formation simulations have utilised these
observations to provide additional constraints on feedback mechanisms
and other physical processes of galaxy
formation~\citep[e.g.][]{Vogelsberger-14,Schaye-15,Dave-16}.  For
instance, the slope of the mass-metallicity relation strongly
suggests that low-mass galaxies preferentially eject more of their
gas in outflows versus forming it into stars~\citep[e.g.][]{Finlator-08}.
The high gas fraction in low-mass galaxies is likewise a reflection
of strong outflows that prevents the gas from forming into
stars~\citep[e.g.][]{Dave-11b}.  This broadly agrees with the notion
that low-mass galaxies must have stronger feedback in order to
suppress the faint end of the GSMF~\citep[e.g.][]{Somerville-08,Dave-11b}.
While these trends generally point towards a qualitatively similar
picture~\citep{Somerville-15}, it remains highly challenging for a
single model to quantitively reproduce all the relevant observed
relations across a wide range of mass scales and cosmic epochs.

In this paper we present a further analysis on the suite of
cosmological hydrodynamic simulations of galaxy formation using
{\sc Gizmo}, called the \mufasa\ simulations, introduced in
\citet[hereafter Paper~I]{Dave-16}.  \mufasa\ uses updated state
of the art feedback modules, including two-phase kinetic outflows
with scalings taken from the FIRE simulations~\citep{Muratov-15},
an evolving halo mass-based quenching scheme~\citep{Gabor-15,Mitra-15},
11-element chemical evolution, and molecular gas-based star
formation~\citep{Krumholz-11,Thompson-14}.  We run three volumes, each with
$512^3$ dark matter particles and $512^3$ gas elements, having box
sizes of 50, 25, and $12.5\hmpc$, in order to cover halo masses
from $\sim10^{10}-10^{14}M_\odot$ and stellar masses from $\sim10^7-10^{12}M_\odot$.

In Paper~I we showed that \mufasa\ does an excellent job at reproducing
the observed evolution of the GSMF over most of cosmic time.  Here
we compare \mufasa\ to a wider suite of observations encompassing
galaxy SFRs, gas, and metal content, in order to quantitatively
examine whether a model that accurately reproduces stellar mass
growth can also match these independent properties.  One significant
discrepancy seen in Paper~I was that specific SFRs (sSFRs) at $z\sim
1-2$ were well below observations, even though galaxy growth rates
as measured by GSMF evolution seemed to be in accord with data.
Here we further investigate this issue using SFR and sSFR functions
over cosmic time.  Since \mufasa\ directly tracks H$_2$ within
galaxies using a sub-grid prescription~\citep{Krumholz-11}, we
investigate \ion{H}{i} and H$_2$ contents separately, along with
their evolution.  Galaxy metallicities provide a crucial barometer
for feedback, so we compare our predictions to emerging observations
out to Cosmic Noon.  Simulations naturally predict that deviations
from the mean galaxy scaling relations are correlated, in that
galaxies at a given stellar mass that are high in SFR are also low
in metallicity~\citep{Dave-11b} and gas content~\citep{Rafieferantsoa-15}.
Here we generalize this analysis across all quantites considered,
showing that deviations from the mean relations in SFR, metallicity,
\ion{H}{i}, and H$_2$ versus $M_*$ are all correlated, and we
quantify these correlations.

Taken together, these results extend the overall success of the
\mufasa\ simulations as a reasonably faithful reproduction of the
real universe, thereby highlighting \mufasa's utility as a platform
to study of the physics of galaxy evolution across cosmic time.
This paper is outlined as follows:  In \S\ref{sec:code} we briefly
recap the key ingredients of the \mufasa\ simulations.  \S\ref{sec:sfr}
discusses predicted SFRs and sSFRs, \S\ref{sec:mzr} presents the
mass-metallicity relation, and \S\ref{sec:fgas} shows gas fractions
and gas mass functions.  In \S\ref{sec:dev} we quantify the
second-parameter dependences of the scatter around key scaling
relations.  We summarize our findings in \S\ref{sec:summary}.

\section{Simulation Description}\label{sec:code}

We employ a modified version of the gravity plus hydrodynamics
solver {\sc Gizmo}~\citep{Hopkins-15a}, which uses the \gad\ gravity
solver~\citep{Springel-05}, along with the meshless finite mass
(MFM) hydrodynamics solver.  We use adaptive gravitational softening
throughout for all particles~\citep{Hopkins-15a}, with a minimum
(Plummer-equivalent) softening length set to 0.5\% of the mean
interparticle spacing.  For more details on these aspects as well
as the feedback choices summarised below, see Paper~I.

We include radiative cooling from primordial (non-equilibrium
ionisation) and heavy elements (equilibrium ionisation) using the
{\sc Grackle-2.1} chemistry and cooling library~\citep{Enzo-14,Kim-14}.
A spatially-uniform photo-ionising background is assumed, namely
the 2011 update of the determination in \citet{Faucher-09}.  Gas
above a threshold density is assumed to have an equation of state
given by $T\propto\rho^{1/3}$~\citep{Schaye-08}, and for the primary
run employed in this paper the threshold density is taken to be
$0.13$~cm$^{-3}$.  Stars are formed using a molecular gas-based
prescription following \citet{Krumholz-09}, which approximates the
H$_2$ fraction based on the local density, the Sobolev approximation
in which the optical depth is given by $\rho/|\nabla\rho|$ where
$\rho$ is the particle's density, and the particle's metallicity
scaled to solar abundance based on \citet{Asplund-09}.  We vary the
assumed clumping factor with resolution, as described in Paper~I.

Young stellar feedback is modeled using decoupled, two-phase winds.
Winds are ejected stochastically, with a probability that is $\eta$
times the star formation rate probability.  The formula for $\eta$
is taken to be the best-fit relation from the Feedback In Realistic
Environments (FIRE) suite of zoom simulations~\citet{Muratov-15}, namely
\begin{equation} \label{eq:eta}
\eta=3.55 \Bigl(\frac{M_*}{10^{10}M_\odot}\Bigr)^{-0.351},
\end{equation}
where $M_*$ is the galaxy stellar mass determined using an
on-the-fly friends-of-friends galaxy finder.  The ejection
velocity $v_w$ scaling is also taken to follow that predicted
by FIRE, but with a somewhat higher amplitude:
\begin{equation}\label{eq:vw}
v_w = 2 \Bigl({v_c\over 200\kms}\Bigr)^{0.12} v_c + \Delta v_{0.25}.
\end{equation}
where $v_c$ is the galaxy circular velocity estimated from the
friends-of-friends baryonic mass, and $\Delta v_{0.25}$ accounts
for the potential difference between the launch location and
one-quarter of the virial radius where \citet{Muratov-15} measured
the scalings from FIRE.  Winds are also ejected with a random 30\% fraction
being ``hot", namely at a temperature set by the difference between
the supernova energy and the wind launch energy (if this is positive), with 
the remaining 70\% launched at $\ll 10^4$K.  Wind fluid elements are allowed to
travel without hydrodynamic forces or cooling until such time as
its relative velocity versus surrounding (non-wind) gas is less
than 50\% of the local sound speed, or alternatively if it reaches
limits in density of 0.01 times the critical density for star
formation, or a time given by 2\% of the Hubble time at launch.  
We further include energy 
Type Ia supernovae (SNIa) and asymptotic giant branch (AGB) stars,
implemented as a delayed component using stellar evolution as
tracked by \citet{Bruzual-03} models with a \citet{Chabrier-03} initial
mass function (IMF).  See Paper~I for full 
details.

Chemistry is tracked for hydrogen, helium, and 9 metals:  C, N, O,
Ne, Mg, Si, S, Ca, and Fe, comprising over 90\% of metal mass in
the universe.  Type~II SN yields are taken from \citet{Nomoto-06},
parameterised as a function of metallicity, which we multiply by
0.5 in order to more closely match observed galaxy metallicities.
Type~II yields are added instantaneously to every star-forming gas
particles at every timestep, based on its current star formation
rate.  For SNIa yields, we employ the yields from \citet{Iwamoto-99},
assuming each SNIa yields $1.4 M_\odot$ of metals.  For AGB stars,
we employ enrichment as a function of age and metallicity from
various sources as described in \citet{Oppenheimer-08}, further
assuming a 36\% helium fraction and a nitrogen yield of 0.00118.
The enrichment, like the energy, is added from stars to the nearest
16 gas particles, kernel-weighted, following the mass loss rate as
computed assuming a \citet{Chabrier-03} IMF.

We note that ISM gas ejected from our simulated galaxies is done
so without any modification to its metallicity.  We do not employ
a separate ``metal loading factor" parameter (i.e. the metallicity
of the ejected gas relative to the ISM metallicity) which preferentially
ejects enriched ~\citep[or de-enriched, as in
Illustris;][]{Vogelsberger-14} ISM material; in other words, we
assume a metal loading factor of unity.  The physical justification
for this is that, particularly in low mass galaxies where the mass
loading factor $\eta$ is high, direct supernovae ejectae represent
only a very small portion of the total outflowing material, hence
it makes sense that the outflow metallicity is dominated by ambient
ISM gas (surrounding the launch site).  In higher-mass galaxies
where $\eta$ is low, this assumption can break down, and it may be
more appropriate to include a metal loading factor greater than
unity.  Without more detailed modeling, it is difficult to determine
exactly what the appropriate metal loading factor is, so we eschew
this complication for the present.  Note that the FIRE simulations
with self-consistently generated outflows find metal loading factors
around unity for all galaxies~\citep{Ma-16}, supporting our assumption.

To quench massive galaxies, we employ an on-the-fly halo mass-based
quenching scheme that follows \citet{Gabor-12,Gabor-15}.  Above a
halo quenching mass $M_q$, we maintain all halo gas at a temperature
above the system virial temperature, by continuously adding heat.
This is intended to mimic the effects of ``radio mode" or ``jet
mode" quenching \citep{Croton-06}, where jets inflate superbubbles
in surrounding hot gas which approximately sphericalises the jet
energy and counteracts gas cooling~\citep{McNamara-07}.  We only
add heat to gas that is not self-shielded, defined as having a
neutral (atomic+molecular) fraction above 10\% after applying a
self-shielding correction following \citet{Rahmati-13}.  We take
$M_q$ as determined from the analytic ``equilibrium model" constraints
required to match the observed evolution of the galaxy population
from $z=0-2$~\citep{Mitra-15}, namely:
\begin{equation}
M_q = (0.96 + 0.48z)\times 10^{12}M_\odot.
\end{equation}
As demonstrated in Paper~I this evolving quenching mass
is nicely consistent with observations during early epochs ($z\sim 2$)
and today, while providing a sharp turnover in the stellar mass function
at late epochs that closely matches observations.

Paper~I focused on the $50\hmpc$ \mufasa\ simulation using $512^3$
gas fluid elements (i.e. mass-conserving cells), $512^3$ dark matter
particles, and $0.5\hkpc$ minimum softening length.  Table~I of
Paper~I lists the details for two higher-resolution runs with the
identical input physics and number of particles, having box sizes
of $25\hmpc$ and $12.5\hmpc$ and proportionally smaller softening
lengths.  At that time, these simulations were only evolved to
$z=2$, but since then we have evolved the $25\hmpc$ volume to $z=0$
and the $12.5\hmpc$ run to $z=1$.  We will use these to extend the
dynamic range of our predictions and to test resolution convergence.

We generate initial conditions at $z=249$ using {\sc Music}~\citep{Hahn-11}
assuming a cosmology consistent with \citet{Planck-15} ``full
likelihood" constraints: $\Omega_m=0.3$, $\Omega_\Lambda=0.7$,
$\Omega_b=0.048$, $H_0=68\kmsmpc$, $\sigma_8=0.82$, and $n_s=0.97$.
We output 135 snapshots down to $z=0$ (105 to $z=1$).  We analyse
the snapshots using SPHGR-yt\footnote{\tt
http://sphgr.readthedocs.org/en/latest/} \citep{Thompson-15}, which
identifies galaxies using SKID and halos using {\sc
RockStar}~\citep{Behroozi-13}, links them via their positions, and
outputs a catalog of properties required for all the analyses in
this paper.

\section{Star Formation Rates}\label{sec:sfr}

Paper~I compared \mufasa\ to the evolution of the stellar mass
function, showing general agreement with the growth of the stellar
content of galaxies across much of cosmic time.  However, it also
reiterated a longstanding discrepancy in predictions of sSFRs at a
given $M_*$, i.e. the main sequence, during the peak epoch of cosmic
star formation, in which simulated galaxies have $\sim\times 2-3$
lower SFRs compared to observations at $z\sim 2$.  Here we explore
the distribution of SFRs in more detail, by comparing \mufasa\ to
two other SFR observables, namely the star formation rate function
and the specific star formation rate function.

\subsection{Star formation rate function}

\begin{figure}
  \centering
  \subfloat{\includegraphics[width=0.47\textwidth]{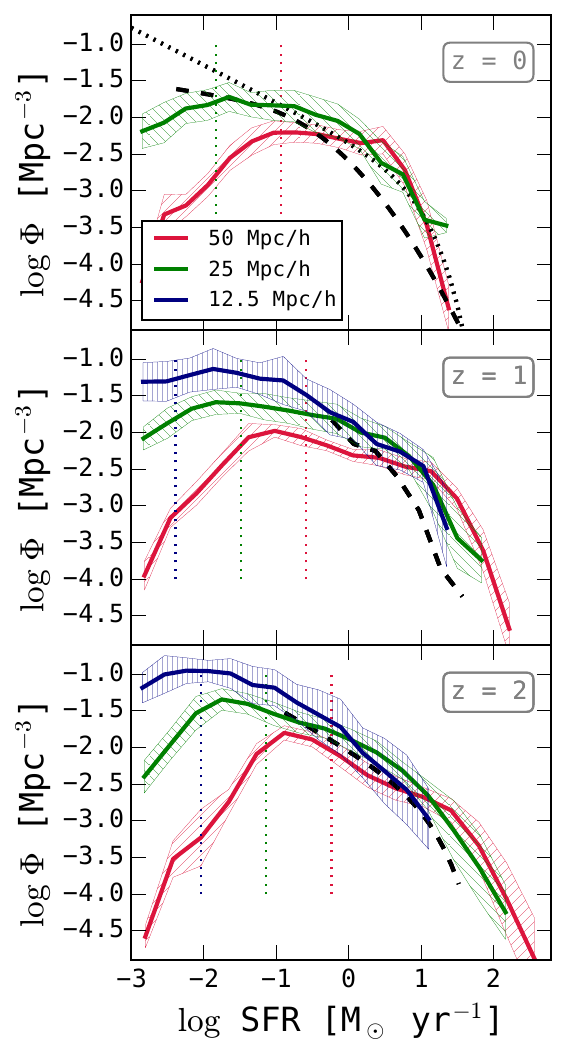}}
  \vskip-0.1in
  \caption{Star formation rate functions at $z=0,1,2$ (top to bottom),
in our suite of \mufasa\ simulations.  Results for the $50\hmpc$, $25\hmpc$, and $12.5\hmpc$ (at
$z\geq1$) are shown in red solid, green dashed, and blue dotted lines, respectively, 
with the hatched
region showing the cosmic variance computed over the 8 sub-octants of the simulation volume.
The vertical dotted line is an approximate resolution limit, taken as the
mean SFR at our stellar mass resolution limit; there is likely some incompleteness
even above this value, which can be seen by comparing the different volumes.
Observations are shown as dashed black lines in each panel, using H$\alpha$
luminosity functions converted to SFRFs from \citet[$z\approx 0$]{Gunawardhana-13},
\citet[$z\approx 1$]{Colbert-13}, and \citet[$z\approx 2$]{Mehta-16}.
}
\label{fig:sfrfunc}
\end{figure}

Figure~\ref{fig:sfrfunc} shows SFR functions (SFRFs) at $z=0,1,2$
from our suite of \mufasa\ simulations.  The red solid, green dashed, and blue dotted
curves show the results from our $50\hmpc$, $25\hmpc$, and $12.5\hmpc$
(at $z\geq1$) simulations.  The hatched region shows cosmic variance
as computed over the 8 sub-octants within each simulation volume.
The vertical dotted line indicates the typical SFR at the steller
mass resolution limit of 32 gas particle masses from a fit to the $M_*-$SFR relation; below
this, the distribution of SFRs is expected to be significantly
compromised by numerical resolution, and even above this SFR there
may be some galaxies that are impacted by poor resolution owing to
the scatter in the $M_*-$SFR relation.  Hence this line should be
regarded as an approximate rather than a strict resolution limit.
Indeed, one can see from comparing the various simulations' SFR
functions at the same SFR that the lack of resolution convergence
seems to begin significantly above the dotted line.

Observations are shown in the various panels from $H\alpha$ luminosity
functions, converted to SFR using the relation taken from
\citet{Kennicutt-98}, adjusted for a Chabrier IMF.  At $z\sim 0$,
we show data from \citet[dotted black]{Bothwell-11} and \citet[dashed
black]{Gunawardhana-13}, at $z\sim 1$ from \citet{Colbert-13}, and
at $z\sim 2$ from \citet{Mehta-16}.  All these observations account
for extinction based on considering H$\beta$ and sometimes more,
but there is still uncertainty in such corrections.

At $z=0$, the simulated SFRF are in good agreement with
\citet{Bothwell-11}, but overpredict by up to $\sim\times 3$ the
more recent \citet{Gunawardhana-13} data from the Galaxy and Mass
Assembly (GAMA) survey.  This implies that there are several times
more SFGs with SFR$\sim 1-10 M_\odot$yr$^{-1}$ in \mufasa\ than in
the real Universe.  It is possible that H$\alpha$ surveys miss the
most highly star-forming galaxies since they are typically highly
obscured.  This could be mitigated by examining far-IR based SFR
estimators, but that introduces the additional complexity of
subtracting off the AGN contribution to the total flux, which is
often substantial in luminous IR galaxies.  This discrepancy is
consistent with the finding in Paper~I (see their Figure~3) that
the cosmic SFR density is overpredicted by $\sim 50\%$ at $z=0$ in
\mufasa.  Hence \mufasa's predictions for the SFRF today are broadly
consistent with data, but with a notable overprediction of galaxies
with SFRs comparable to or exceeding that of the Milky Way.

At $z=1$, the predicted SFRF is similar to that at $z=0$ at low
SFRs, but shows an excess at high SFRs, such that now we start to
see galaxies with SFR$\ga 100M_\odot$yr$^{-1}$ in our $50\hmpc$
volume.  The SFRF does not show as strong a truncation at high-SFR
as it does at low-$z$.  Generally, \mufasa\ exceeds observations at high SFRs,
albeit with the same caveats regarding highly obscured galaxies
that become more prevalent at high $z$.

At $z=2$, the trend continues that the low-SFR end is mostly
unevolving but the high-SFR end is more highly populated.  Once
again there is an excess in \mufasa\ relative to data, but it is
fairly mild at this epoch.  Interestingly, although \mufasa\ seems
to reproduce the GSMF well at this epoch, and if anything the SFRF
is overpredicted, it nonetheless yields an SFR$-M_*$ relation that
is clearly too low~(Paper~I).  Notably, the main sequence is typically
derived from UV and/or rest-near infrared measures of SFR, not
H$\alpha$.  It is possible that various systematics operate differently
at this epoch among the various observational SFR indicators.  It
is beyond the scope of this work to fully examine all the relevant
systematics, but it highlights that, leaving aside the models, there
appears to be some consistency issues purely among observational
measures of SFRs during Cosmic Noon.

In summary, the SFRFs predicted by \mufasa\ generally show the
observed shape from $z=0-2$, though with an amplitude that is
somewhat too high at low redshifts.  The broad agreement is encouraging
and may be within current systematic uncertainties in measuring a
complete sample of star-forming galaxies across all these epochs.
There is no obvious discrepancy in the SFRF at $z=2$ that would
explain the discrepancy in the SFR$-M_*$ relation.  Resolution
convergence in the SFRF between the various \mufasa\ volumes is
reasonable, though not ideal.

\subsection{Specific star formation rate function}

\begin{figure*}
  \centering
  \subfloat{\includegraphics[width=0.8\textwidth]{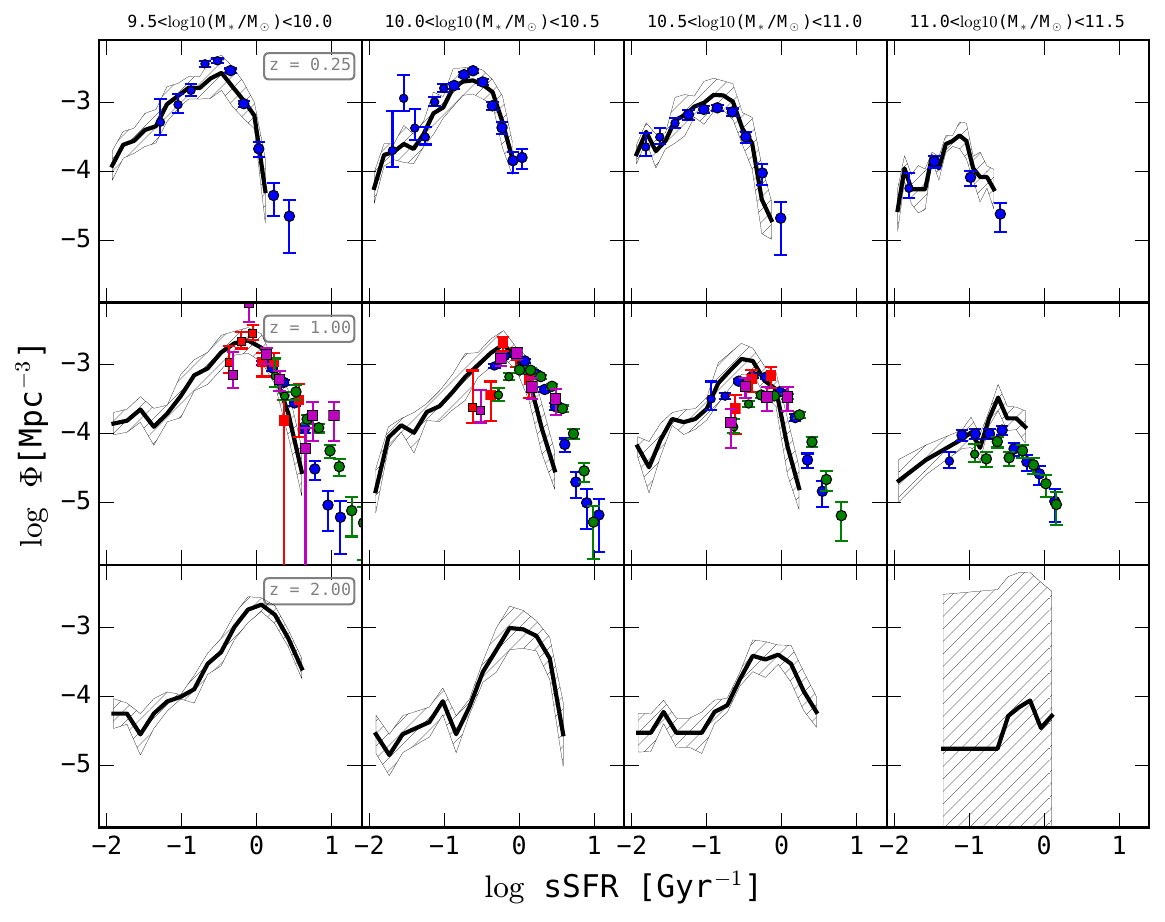}}
  \vskip-0.1in
  \caption{Specific star formation rate functions at $z=0.25,1,2$
(top to bottom rows) in our $50\hmpc$ \mufasa\ simulation, in four bins
of increasing stellar mass (left to right).  Hatched regions
show the cosmic variance computed over the 8 sub-octants in the volume.
Observations from \citet{Ilbert-15}
are shown, which includes data from COSMOS at $z=0.2-0.4$ (blue in upper
panels), and in the middle panels from COSMOS at $z=0.8-1$ (blue), 
COSMOS at $z=1-1.2$ (green), GOODS from $z=0.8-1$ (red), and GOODS from
$z=1-1.2$ (magenta).  The predicted sSFR functions match observations very
well at $z\sim 0.25$, showing that \mufasa\ reproduces the distribution
of sSFRs quite well at low-z.  At $z=1$ \mufasa\ matches well the shape
of the distribution but is shifted to slightly lower sSFR.  This indicates
that \mufasa\ is properly capturing the physical causes of fluctuations
around the main sequence, as well as the number of galaxies transitioning
to quiescence.
}
\label{fig:ssfrdist}
\end{figure*}

A separate test of SFRs is whether our simulations reproduce the
correct distribution of specific star formation rates at a given
stellar mass.  Qualitatively, at high redshifts the spread in sSFRs
measures the fluctuations around the main sequence owing to inflow
fluctuations~\citep[e.g.][]{Mitra-16}, while at lower redshifts a
substantial low-sSFR population appears corresponding to quenched
galaxies.  Matching the amplitude and evolution of the distribution
of sSFRs in stellar mass bins is thus a stringent test of whether
the predicted \mufasa\ galaxy population is in accord with the rate
at which galaxies are fluctuating around the main sequence, and
eventually quenched~\citep[e.g.][]{Tacchella-16}.

Figure~\ref{fig:ssfrdist} shows the specific star formation rate
function (sSFRF) in four bins of stellar mass from
$10^{9.5}<M_*<10^{11.5}M_\odot$ (left to right), at $z=0.25, 1, 2$
(top to bottom).  We only consider the $50\hmpc$ here volume for
clarity, particularly since we want to well sample the rate of
galaxy quenching for which we prefer our largest volume containing
the most massive halos.  Lines show the predicted sSFRF, while the
hatched region shows the cosmic variance computed among 8 sub-octants.
Observations are shown from a compilation by \citet{Ilbert-15} at
$z=0.2-0.4$ and $z=0.8-1.2$, from various sources as described in
the caption, generally from extinction-corrected UV measures or SED
fitting.  Note that the observations only consider galaxies which
have a measurable SFR, which we mimic in our simulations by excluding
galaxies with $\log$~sSFR$<-3$ (which would lie off this plot in
any case).

At $z=0.25$, the sSFRF shows a peak at the median sSFR within that
$M_*$ bin, a sharp truncation to higher sSFR, and a broader extension
to low sSFR corresponding to green valley galaxies.  \mufasa\
provides a remarkably good match (i.e. within cosmic variance) to
the observed sSFRF in every stellar mass bin.  This new test of
models demonstrates that the scatter in sSFRs, and hence the
fluctuations around the main sequence as well as the rate at which
the green valley is being populated, is being well modeled in
\mufasa.  In particular, the amplitude and shape match in the most
massive bin would suggest that \mufasa\ is not overproducing the
number of galaxies with high sSFRs, even if Figure~\ref{fig:sfrfunc}
suggested that it might be doing so.  These can be reconciled if
\mufasa\ is producing a few too many massive galaxies, which is
indeed a trend noted in the $z=0$ GSMF shown in Paper~I, albeit
with large cosmic variance.

At $z=1$, the shape of the sSFRF is well reproduced, but there is
clearly an offset in the distribution such that the predicted values
are lower by $\sim\times 2$.  This is simply reflecting the fact
that the median sSFR is underproduced at this epoch, as shown in
Paper~I, continuing a trend generically seen in
cosmological galaxy formation models.  It appears that the discrepancy
in the median sSFR is not reflective of the emergence of some new
population of galaxies in observations that do not appear in the
models, but rather an overall systematic shift in the measured sSFR
values at that epoch.  We would expect that these trends would
continue on to $z=2$, but we do not know of sSFRFs published at
this epoch.

Overall, \mufasa\ does an excellent job of reproducing the low-$z$
distribution of sSFRs, including the peak value, the sharp truncation
to high sSFRs that highlights the rarity of starbursts locally, and
the gradual decline towards low-sSFR that reflects the population
of galaxies likely in the process of quenching.  There are still a
non-trivial number of SFGs even at the highest masses in \mufasa,
which is in agreement with observations.  This suggests that \mufasa\
does a good job reproducing the SFR fluctuations and quenching rate
of galaxies, which provides some empirical support for the implemented
subgrid models for star formation and quenching.

\section{Metallicity}\label{sec:mzr}

Chemical enrichment provides a key tracer for star formation and
feedback activity in and around galaxies.  Within a simple equilibrium
or bathtub-type model, the mass-metallicity relation directly
reflects the mass loss rate in outflows together with the recycling
of previously-ejected (enriched) material back into the
ISM~\citep[e.g.][]{Finlator-08,Somerville-15}.  Galaxy metallicities
are thus a crucial test for how accurately a particular model is
representing the baryon cycle.

The stellar mass--gas phase metallicity relation (MZR) is one of
the tightest observed correlation between any two galaxy properties,
with a scatter typically around 0.1~dex~\citep{Tremonti-04}.
Unfortunately, calibration issues may add significant systematic
uncertainties~\citep{Kewley-08}, but nonetheless the shape of the
MZR is likely to be reasonably robust even if the amplitude is less
certain.  In this section we present predictions for the MZR from
\mufasa, along with comparisons to key observations at the present
epoch and in the early Universe.

\begin{figure*}
  \centering
  \subfloat{\includegraphics[width=0.5\textwidth]{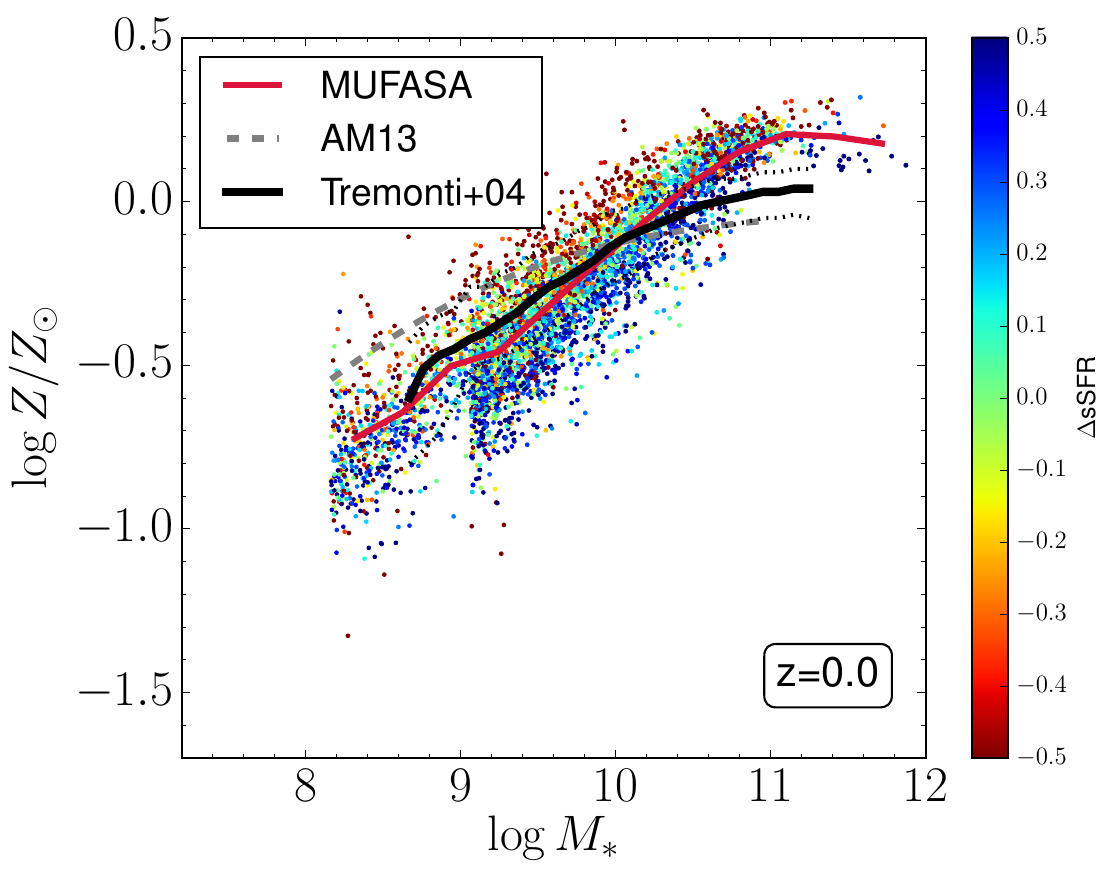}}
  \subfloat{\includegraphics[width=0.5\textwidth]{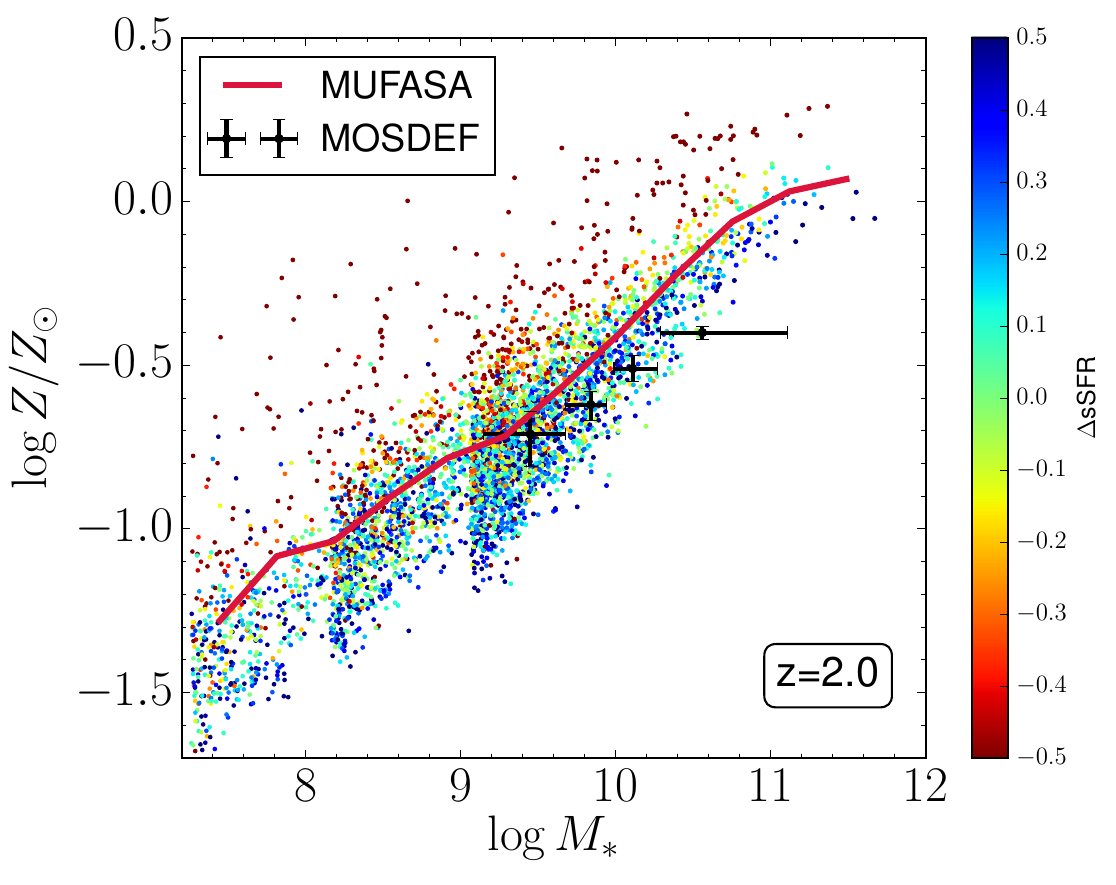}}
  \vskip-0.1in
  \caption{\mufasa\ galaxy mass-metallicity relations at $z=0,2$,
computed from the predicted SFR-weighted oxygen abundance assuming the solar
oxygen abundance from \citet{Vagnozzi-16}.  Displayed points are combined
from the $50\hmpc$, $25\hmpc$, and $12.5\hmpc$
runs, and for each run every galaxy with gas is plotted down to the 64-particle 
stellar mass resolution limit where a break is evident.
Points are colour-coded by their distance from the $M_*-$SFR
relation; bluer points have higher SFR for their $M_*$, as indicated
by the colour bar.  Observations at $z=0$ are shown from \citet[T04;
solid black]{Tremonti-04} and \citet[AM13; grey dashed]{Andrews-13},
while $z\approx 2$ data is shown from \citet{Sanders-15}.
}
\label{fig:mzr}
\end{figure*}

Figure~\ref{fig:mzr} shows the MZR at $z=0, 2$ (left, right panels)
in our \mufasa\ simulation suite.  At $z=2$, we have overplotted
all three volumes down to each of their galaxy stellar mass resolution
limit; these are the three ``groupings" of points, with the $12.5\hmpc$
volume extending to the lowest masses, and the $50\hmpc$ volume
dominating at high masses.  At $z=0$, we only have the $50\hmpc$
and $25\hmpc$ volumes.  The thick red line shows a running median
for the combined sample of simulated galaxies; while we do not show
the individual volumes' medians separately, it is evident that the
agreement between them is reasonable in the overlapping mass ranges,
as there is no significant break in the median fit when crossing
over a mass resolution threshold, though higher-resolution simulations
tend to predict slightly higher metallicities at a given mass.  The
colour coding shows the deviation in $\log$~SFR for each galaxy off
of the global $M_*-$SFR relation at that redshift (Paper~I).
Observations at $z=0$ are shown from the Sloan Digital Sky Survey
(SDSS), via nebular line fitting~\citep[black solid line]{Tremonti-04}
and ``direct" abundance measures from stacked spectra~\citep[grey
dashed line]{Andrews-13}.  At $z=2$, we show observations from the
Mosfire Deep Evolutionary Field (MOSDEF) survey~\citep{Kriek-15}
using O3N2 abundances obtained from near-infrared Keck
spectroscopy~\citep[points with errorbars;]{Sanders-15}.

Broadly, the agreement between \mufasa\ and observations is fairly
good.  The faint-end slope is generally consistent with data at
both redshifts, and at high-$z$ it can be seen that the simulated
MZR slope extends unabated to much lower masses than can be observed
prior to the {\it James Webb Space Telescope}.  At the massive end,
there is clearly a turnover at low redshifts above $M_*\ga
10^{11}M_\odot$, and even at $z=2$ there is a hint of a similar
turnover though even the $50\hmpc$ volume does not adequately probe
the very high-mass end at that epoch.

At low masses, there is $\sim 0.2-0.3$~dex increase in the metallicity
at a fixed $M_*$ from $z=2\rightarrow 0$.  The evolution is slightly
less at high masses, creating a more prominent flat portion of the
MZR.  This amount of evolution, and the trend of a more prominent
turnover at low masses, is generally consistent with
observations~\citep{Zahid-14,Steidel-14,Sanders-15}.

A more careful comparison to MZR data reveals some notable
discrepancies.  Most obviously, there is a clear overprediction of
the metallicity at $M_*\ga 10^{10.3}M_\odot$ at $z=0$.  It appears
that the high-mass flattening begins at a lower mass scale in the
data as compared to in \mufasa, which continues with an unabated
power-law up to nearly $10^{11}M_\odot$ before flattening.  There
is even a hint of such an overproduction at $z=2$; while the overall
amplitude is slightly too large compared to these observations at
all masses, this is particularly exacerbated for the highest mass
bin.  One possibility for reconciling this in the models would be
that the metal loading factor at $M_*\ga 10^{10.3}M_\odot$ should
be greater than unity, which would preferentially eject a higher
fraction of metals out of high mass galaxies.  Alternatively, it could
be that the models have excess wind recycling at high masses; we
will examine mass flows and recycling in detail in future work.

One can also see that the low-mass end of the MZR is in better
agreement with the \citet{Tremonti-04} nebular line MZR than the
direct abundances measures by \citet{Andrews-13}.  Such discrepancies
between observational analyses highlight the difficulty in robustly
calibrating metallicity indicators~\citep{Kewley-08}.  Moreover,
at high redshifts it is possible that the typical stellar population
in $z\sim 2$ star-forming galaxies may be substantially different
than that at low redshifts~\citep{Steidel-16}, which could alter
the usual metallicity calibrations applied to nebular emission line
measures.  In light of this, the disagreements between \mufasa\ MZR
predictions and observed may be regarded as preliminary.

Finally, the colours of the points show a clear trend that galaxies
with low sSFR at a given mass will have high metallicity, and vice
versa.  This has been noted in
data~\citep{Ellison-08,Lara-Lopez-10,Mannucci-10,Salim-14,Telford-16}, and \citet{Mannucci-10}
dubbed this the fundamental metallicity relation (FMR) because they
further argued that the SFR$-M_*-Z$ relation was also redshift-independent.
More recent results have called into question whether the FMR is
truly redshift independent~\citep{Salim-15,Brown-16,Grasshorn-16},
and also whether it is even seen at high
redshift~\citep{Steidel-14,Sanders-15}.  However, it appears that
the samples at $z\sim 2$ may not be sufficient for such a trend to
have been apparent, and moreover calibration issues can mask such
subtle correlations~\citep{Salim-15}.  It is thus unclear whether
the FMR exists at $z\sim 2$ observationally.  We will discuss this
second-parameter dependence of the MZR on the sSFR further in
\S\ref{sec:dev}.

In \mufasa, the general trend of the SFR$-M_*-Z$ relation is apparent
at both $z=0$ and $z=2$.  However, the predicted MZR is notably
tighter at $z=0$ (typical variance of $\sigma\approx 0.1$ dex around
the mean relation) than at $z=2$ ($\sigma\approx 0.2$ dex).  By
$z=2$, the most metal-rich galaxies already have metallicities
comparable to the most metal-rich objects at $z=0$, across all
$M_*$, while the most metal-poor objects are much less enriched.

The physical explanation for the second-parameter correlation with
SFR is that an increase in gas accretion will bring in metal-poor
gas while fueling new star formation, and conversely a lull in
accretion will result in an evolution more similar to a closed box
that will raise the metallicity quickly by consuming its
gas~\citep[e.g.][]{Finlator-08}.  As pointed out in \citet{Dave-11b},
the lull is permanent for satellite galaxies, causing them to 
reach a slightly higher metallicity at a given mass before running
out of fuel, as observed~\citep{Pasquali-10}; though we don't show
it here, this is true in \mufasa\ as well.  Hence in the fluctuating
``smooth accretion" scenario for galaxy fueling~\citep{Keres-05,Dekel-09},
the FMR is a natural outcome, and the scatter about the relation
reflects the frequency and impact of accretion flucutations such
as mergers.  Confirming the reality of the FMR at $z\sim
2$ is thus a crucial test of this scenario.  In \S\ref{sec:dev} we
will quantify predictions for this second-parameter correlation
that can be tested against present and future observations.

\begin{figure}
  \centering
  \subfloat{\includegraphics[width=0.45\textwidth]{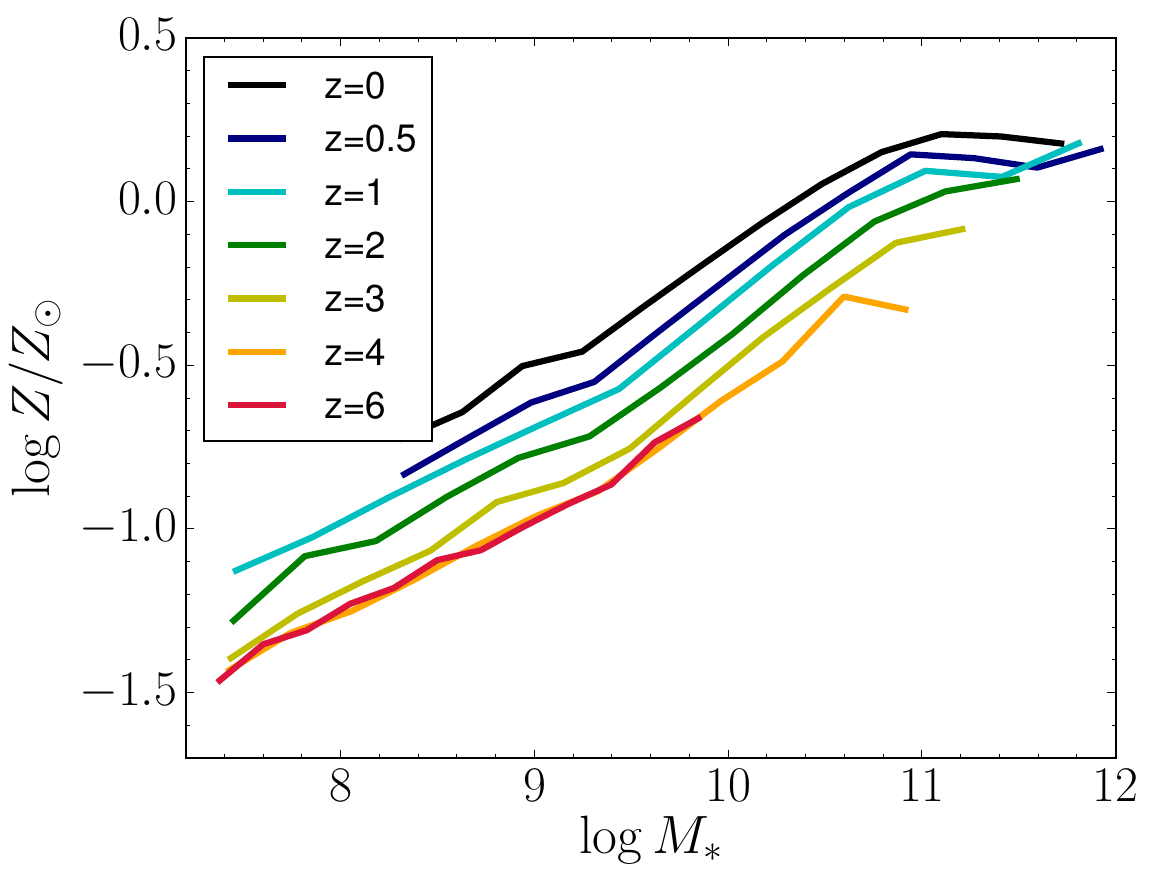}}
  \vskip-0.1in
  \caption{\mufasa\ median galaxy mass-metallicity relations at $z=0,0.5,1,2,3,4,6$.
Displayed relations combine galaxies from the $50\hmpc$, $25\hmpc$, and $12.5\hmpc$
runs down to each of their 64-particle resolution limit.  There is steady upwards
evolution of the MZR over time since $z\sim 4$, with $\sim\times 2$
increase in metallicity at a given $M_*$ since $z=2$.
}
\label{fig:mzrevol}
\end{figure}

Figure~\ref{fig:mzrevol} shows the evolution of the MZR from
$z=6\rightarrow 0$, computed as a running median from the combined
sample of 3 runs.  The MZR shows a constant low-$M_*$ power-law
slope of $\approx 0.5$ at all redshifts.  At the lowest redshifts,
there is the onset of a flattening in the MZR at $M_*\ga
10^{10.7}M_\odot$.  The MZR rises steadily but slowly with time.
At a given mass (below the flattening), the evolution is approximately
0.2~dex out to $z\sim 1$, and then 0.1~dex per unit redshift out
to $z\sim 4$, and no further evolution to $z=6$.  In \citet{Finlator-08}
it was argued that, barring any evolution in $\eta$ with $M_*$
(\mufasa\ assumes none), then the evolution of the MZR must reflect
the enrichment level of accreted material, i.e. wind recycling.  It
remains to be seen if such a scenario is consistent with a
mass-independent increase in the metallicity down to quite low
masses.

Qualitatively, the generally slow evolution and the onset of a
high-mass flattened portion at lower redshifts is consistent with
observations~\citep[e.g.][]{Zahid-14}, as well as data-constrained
analytic models of galaxy evolution~\citep{Mitra-15}.  However, the
mass at which the flattening occurs is generally much higher in
\mufasa\ than in such data, where the onset of flattening is typically
below $M_*\la 10^{10}M_\odot$.  This again reflects the fact that
\mufasa\ appears to produce too steep an MZR at $10^{10}\la M_*\la
10^{11}M_\odot$.

Overall, the slope and evolution of the MZR is in broad agreement
with observations, showing mild evolution out to $z\sim 4$.  However,
there is a key discrepancy around $L^*$ galaxies that bears further
investigation.  In future work we will examine the detailed origin
for the evolution of the MZR, highlighting contributions from in
situ enrichment versus pre-enriched accreted gas.

\section{Gas Content}\label{sec:fgas}

The gas content of galaxies provides a measure of the fuel available
for new star formation.  Molecular gas (H$_2$) directly traces
material that is forming into stars, while atomic gas (\ion{H}{i})
typically resides in a more extended reservoir that connects the
ionised IGM with the molecular ISM.  Hence the gas content of
galaxies represents a combination of the effects of how gas is
converted into stars within the ISM, as well as the processes that
fuel new star formation via gas from the IGM.

Observationally, it is generally believed that the atomic gas in
galaxies evolves slowly out to high redshifts, while molecular gas
evolves more rapidly upwards.  The canonical explanation for this
is that \ion{H}{i} represents a transient reservoir which does not
directly trace star formation, while H$_2$ traces star-forming gas
much more closely and hence drops with time in a manner similar to
what is seen for the cosmic star formation rate.  

In actuality, the story is more subtle.  In simple terms one can
rewrite the ratio of star-forming gas to stars as
\begin{equation}\label{eqn:mgas}
{M_{\rm gas}\over M_*} = {M_{\rm gas}\over SFR}{SFR\over M_*} = t_{\rm dep} {\rm sSFR},
\end{equation}
where the first term is the depletion time and the second term is
the specific SFR~\citep[e.g.][]{Dave-12}.  Given a fixed depletion
time, one then expects the gas content of high redshift galaxies
to be increased.  However, one also expects the depletion time to
be reduced to higher redshifts, since galaxies typically form a
relatively fixed fraction of their gas into stars per dynamical
time~\citep{Kennicutt-98}, and disk dynamical times are expected
to scale approximately with the Hubble time~\citep{Mo-98}.  
If sSFR$\propto (1+z)^{2.5}$, and $t_{\rm dyn}\propto H^{-1}(z)$, 
then one gets approximately $f_{H2}\propto (1+z)$.  Hence galaxies
are expected to have higher star-forming gas fractions at earlier epochs.

Meanwhile, the evolution of atomic hydrogen is not so straightforward
to predict.  In the simplest model where the timescale to pass
through the atomic phase also scales with the halo (or, equivalently,
disk) dynamical time, \ion{H}{i} should follow H$_2$.  But physically,
atomic gas occurs when gas can self-shield against ionising
radiation, yet is not dense enough to be molecular (i.e.  to
self-shield against H$_2$ dissociating radiation).  At high redshifts,
gas is physically denser and accretion is more
filamentary~\citep{Dekel-09}, but the ionising background is stronger.
Which effect wins will depend on the detailed interplay of how gas
is accreted around galaxies.

In this section we examine how the atomic, molecular, and total
neutral (atomic+molecular) gas evolves within galaxies, as a function
of stellar mass, in terms of mass functions, and globally as a
cosmic mass density.

\subsection{Gas fractions}

\begin{figure}
  \centering
  \subfloat{\includegraphics[width=0.47\textwidth]{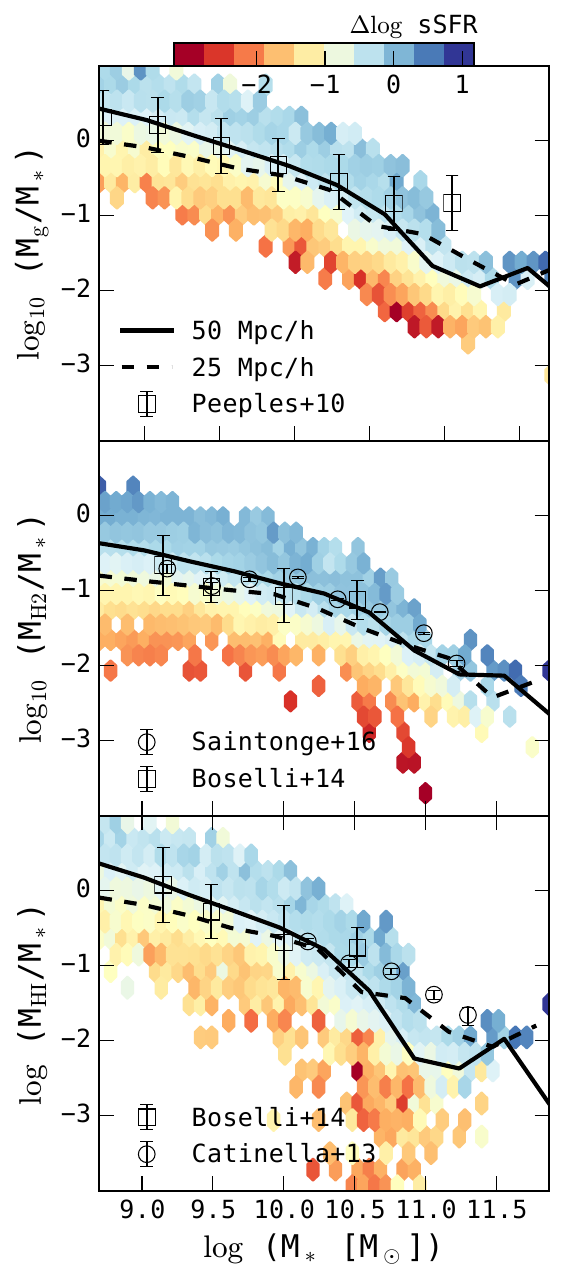}}
  \vskip-0.1in
  \caption{Total neutral (\ion{H}{i}+H$_2$; top panel), molecular (middle),
and atomic (bottom)  gas fractions as a function
of stellar mass predicted from the \mufasa\ $50\hmpc$ simulation
at $z=0$.  Running medians are shown as the solid black lines.
Colour-coding shows the mean sSFR deviation from the main sequence ($\Delta\log$ sSFR) 
in each hexbin.  Dashed line shows the running median from the $25\hmpc$
run to assess resolution convergence.  Data is shown in
the top panel from the compilation by \citet{Peeples-11}, in the middle panel 
from COLDGASS~\citep{Saintonge-16} and HRS~\citep{Boselli-14}, and in the bottom 
panel from HRS and GASS~\citep{Catinella-13}.  The $50\hmpc$
box shows good agreement with observations over the mass range
probed by the data, but the $25\hmpc$ run tends to show lower gas 
fractions at low masses.  At a given $M_*$, galaxies with higher
gas content have higher SFR, and the trend appears tighter for H$_2$.
}
\label{fig:fgas}
\end{figure}

\mufasa, like many recent simulations of galaxy formation, tracks
the amount of molecular gas formed in galaxies.  Owing to limitations
of resolution, this is done via a sub-resolution prescription as
described in Paper~I, broadly following \citet{Krumholz-09} with
minor additions.

Meanwhile, the atomic gas fraction is typically significant only
in regions that are able to self-shield against the cosmic metagalactic
flux (ignoring, as we do here, ionising radiation emitted locally
by the galaxy itself).  Hence we must account for self-shielding
in order to separate the neutral gas from the ionised gas.

We follow the prescription in \citet{Rahmati-13} for determining
the self-shielded fraction.  They provide a fitting formula to the attenuation
in the cosmic metagalactic flux as a function of local density, based on
full radiative transfer simulations.  Given the attenuated ionising flux
impinging on each gas particle, we then compute the rate balance
equations to determine the equilibrium atomic fraction following
\citet{Popping-09}.  For particles at low densities ($n_H\la
10^{-3}$cm$^{-3}$) the gas is generally optically thin, but above
this density one quickly gets more self-shielded gas, increasing
the fraction to unity typically above $n_H\ga 10^{-2}$cm$^{-3}$.
From this self-shielded gas, we then subtract the molecular fraction
as tracked directly in the simulation, which yields the atomic
fraction.  We compute a galaxy's \ion{H}{i} content by summing all
atomic gas that is more gravitationally bound to that galaxy
relative to any other galaxy, using the total baryonic mass to
compute the gravitational binding.  In practice, we do not consider
gas with $n_H<10^{-4}$~cm$^{-3}$ since this is never self-shielded
and thus contributes negligibly to the total \ion{H}{i} content.

Figure~\ref{fig:fgas} shows the total (\ion{H}{i}+H$_2$) (top panel),
molecular (middle), and atomic (bottom) gas fractions as a function
of stellar mass at $z=0$.  The solid black line shows a running median for
the fiducial $50\hmpc$ volume.  The overlaid hexbins are colour-coded by the
average sSFR at that gas fraction relative to the global average
sSFR at the given $M_*$. The dashed line shows a similar running
median for the $25\hmpc$ run, to illustrate the level of resolution
convergence.  

In the top panel, the total gas fraction as a function of $M_*$ in
the $50\hmpc$ run is in excellent agreement with a compilation of
observations by \citet{Peeples-11} over most of the mass range.  At
the highest masses, the observations lie above the model predictions.
While these data only include galaxies where gas was detected, and
many of the simulated galaxies have such low gas fractions that
they would likely evade detection, since there are no predicted
galaxies at all at the median total gas fraction, it appears at
face value that the discrepancy is real.  For $M_*\la 10^{10.5}M_\odot$,
however, galaxy samples are quite complete, and hence the agreement
is a robust success of the models.

The middle panel shows that the molecular gas fractions are likewise
in good agreement with observations from the COLDGASS
survey~\citep{Saintonge-16}, as well as the Herschel Reference
Survey~\citep[HRS;][]{Boselli-14}.  COLDGASS~\citep{Saintonge-11}
is an $M_*$-complete survey and hence is quite directly comparable
to our simulated galaxies.  \mufasa\ even traces the slight turn-down
in $f_{H2}$ at $M_*\ga 10^{10.5}$ relative to an extrapolated trend
from lower masses, which is indicative of a typical mass scale at
which quenching kicks in.

The atomic gas fractions are compared to data from the GASS
survey~\citep{Catinella-10}, which is the parent survey of COLDGASS
and hence also a $M_*$-selected sample of SDSS galaxies down to
very low \ion{H}{i} fractions.  At low masses ($M*\la 10^{10.5}M_\odot$)
there is quite good agreement with the GASS data, which again is a
non-trivial success.  However, our $50\hmpc$ volume predicts a sharp
drop in $f_{HI}$ above this mass, whereas the data show a more
gradual trend.  This is likely the origin of the discrepancy in the
total gas fraction at these masses, since $f_{H2}$ shows good
agreement in this mass range.

The $25\hmpc$ volume (dashed lines) consistently shows lower gas
content at $M_*\la 10^{10}M_\odot$, and thus a shallower trend with
$M_*$ that results in a $\sim\times 2$ deficit with respect to the
$50\hmpc$ volume at the lowest probed masses.  The deficit is
essentially identical in both \ion{H}{i} and H$_2$, which suggests
that gas consumption is more rapid in the $25\hmpc$ volume, likely
owing to its higher resolution that achieves higher densities where
more rapid star formation can occur.  Interestingly, this volume
shows no ``dip" in the \ion{H}{i}, and hence total gas, content at
$M_*\sim 10^{11}M_\odot$, indicating that the disagreement in the
$50\hmpc$ may be a peculiarity in that simulation or else some issue
with resolution convergence in terms of the way it interacts with
the quenching model.  One possibility is that the $25\hmpc$ is able
to self-shield gas in massive halos more effectively owing to its
ability to resolve clumpier structures, and thus the quenching model
is less impactful here since by construction it only operates on
non-self-shielded gas.  In any case, at high and low masses it
appears that resolution convergence is not ideal for predicting gas
fractions, and the resulting systematic uncertainties are of the
order of a factor of two.

The coloured hexbins show that at a given $M_*$, both molecular and
atomic gas content are highly correlated with ongoing star formation.
In both cases, galaxies with enhanced gas content for their $M_*$
also have higher sSFR.  The trend appears to be qualitatively
stronger in the molecular case, which is unsurprising since stars
form out of molecular gas in our simulations.  Nonetheless it is
also clearly present in the atomic gas, indicating that the \ion{H}{i}
reservoir plays a role in regulating star formation even if it is
not directly forming stars.  This is qualitatively consistent with
observations that show more low-metallicity gas in the outskirts
of bluer (i.e. higher sSFR) galaxies~\citep{Moran-12}.

As with the metallicity, the qualitative explanation of this is
that a temporary enhancement (lull) of accretion results in both
increased (decreased) gas content and star formation, along with
the aforementioned reduction (increase) in metallicity.  Since it
takes some time for the inflowing gas to first turn into atomic gas
and then molecular and finally stars, the molecular gas content is
expected to be more highly correlated with the SFR.  Hence as with
the FMR, the second parameter dependence of gas content on SFR most
directly reflects fluctuations in the inflow rate~\citep{Mitra-16};
we will quantify this in \S\ref{sec:dev}.

Other models that track molecular and atomic gas generally predict
\ion{H}{i} and H$_2$ fractions broadly in accord with observations,
be they semi-analytic~\citep{Lagos-11,Popping-14} or state-of-the-art
hydrodynamic models such as EAGLE~\citep{Lagos-15,Crain-16}.  Together
with \mufasa's success, this suggests that the overall gas content
is a fairly robustly predictable quantity in models, at least at
$z=0$.  We note that all these models (including ours) have been
tuned at various levels in order to match the present-day stellar
mass function.  It may be that predicting this correctly, plus
having a molecular gas-based prescription for converting gas into
stars, generically leaves the proper amount of gas in galaxies.  If
so, this represents a non-trivial success for current models of
galaxy formation.

\subsection{Gas fraction evolution}

\begin{figure}
  \centering
  \subfloat{\includegraphics[width=0.42\textwidth]{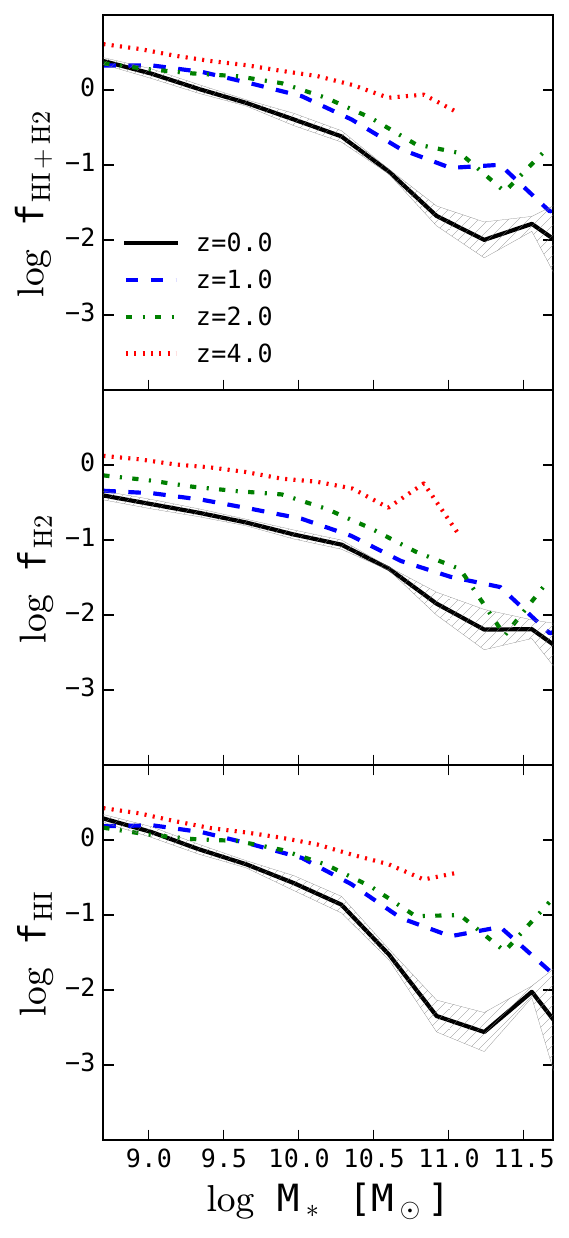}}
  \vskip-0.1in
  \caption{\mufasa\ median gas fractions as a function of $M_*$ 
at $z=0$ (solid black), $z=1$ (dashed blue), $z=2$ (dot-dashed green), and $z=4$ (dotted red)
from the $50\hmpc$ run.
Cosmic variance over 8 sub-octants is shown as the hatched region around the $z=0$ curve.
The \ion{H}{i} fraction (bottom panel) decreases with time
for massive galaxies, while low-mass galaxies are always \ion{H}{i}-dominated.
The molecular gas fraction (middle panel) increases steadily across all masses, but is
typically sub-dominant to \ion{H}{i}.  The total molecular plus atomic content
(top panel) is thus driven by the \ion{H}{i} evolution.
}
\label{fig:fgasevol}
\end{figure}

Galaxies at a given mass are observed to be more molecular gas-rich
at earlier epochs~\citep[e.g.][]{Geach-11,Tacconi-13}.  The amount
of evolution is subject to some uncertainties regarding the conversion
between the observed CO intensity and the molecular gas mass~\citep[see
e.g.][]{Bothwell-13}, but this is unlikely to erase the qualitative
trend.  Far-IR dust continuum measures can also be used to probe gas
content evolution, though likewise subject to some uncertainties regarding
the conversion of dust to gas mass~\citep[e.g.][]{Scoville-16}; in general,
such studies tend to find less strong evolution than CO-based studies.

Neutral gas above $z\ga 0.3$ is currently only observable in
absorption line studies such as with \ion{Mg}{ii} absorbers~\citep{Rao-06}
or Damped Lyman Alpha (DLA) systems~\citep{Prochaska-09}; it is not
obvious how these systems trace galaxies, as it is usually challenging
to identify the individual galaxy giving rise to such absorbers, though
clustering measures offer some general guide that they typically live in $10^{11-12}M_\odot$
halos~\citep[e.g.][]{Bouche-05,Font-Ribera-12}.  Measuring
\HI\ fractions requires having a measure of the stellar mass from
optical or near-IR data for individual galaxies, for which 21~cm
emission can be observed.  While current \HI-21~cm surveys only
probe to $z\sim 0.4$~\citep{Fernandez-16}, upcoming radio telescopes
promise to push direct \ion{H}{i}-21~cm gas content measures in
optically-selected samples out to $z\ga 1$, for example using the
new MeerKAT telescope~\citep{Holwerda-12}, and will be further
advanced with the SKA.  Here we make testable predictions for gas
fractions which can guide such efforts.

Figure~\ref{fig:fgasevol} shows the evolution at $z=0,1,2,4$ of the median total
(\ion{H}{i}+H$_2$) gas fraction versus stellar mass from the $50\hmpc$
\mufasa\ run in the top panel, and the next two panels show this
subdivided into molecular and atomic gas fractions.   The $z=0$
curve is identical to that in Figure~\ref{fig:fgas}, but here we
also show show with shading the cosmic variance estimated via
jackknife resampling over 8 volume sub-octants.  Here we do not
show the second-parameter dependence on SFR as we did in
Figure~\ref{fig:fgas}, but a similar trend persists at all redshifts.
We do not explicitly show any observations on this plot, since
molecular gas observations span some range depending on the type
of data, while atomic gas fraction measures do not yet exist at
$z\ga 0.4$.

The total gas content of galaxies at a given $M_*$ is higher at
earlier epochs.  There appears to be some mild mass dependence to
this statement, as high-mass galaxies lose their gas more quickly
than low-mass galaxies, with an overall effect of steepening the
$f_{\rm gas}-M_*$ relation.  Much of the evolution occurs from
$z\sim 1\rightarrow 0$, prior to which the evolution was somewhat
slower.

Neutral hydrogen (bottom panel) represents the majority of the cold
gas content of galaxies at almost all epochs and masses, except at
high masses today.  Hence the evolutionary trends in \ion{H}{i}
fraction tend to drive those of the total gas content.  The strong
evolution particularly at high masses out to $z\sim 1$ is good news
for upcoming \ion{H}{i} surveys designed to measure 21~cm emission
from galaxies out to this epoch such as LADUMA, and will figure
prominently in the evolution of the \ion{H}{i} mass function discussed
in \S\ref{sec:himf}.

Although we don't show it, the $25\hmpc$ box actually shows quite
good resolution convergence with the $50\hmpc$ box shown here for
all redshifts {\it except} $z=0$.  At $z=0$, the $25\hmpc$ volume
shows a flatter relation (as seen in Figure~\ref{fig:fgas}), but
at higher-$z$ the relations are similar, which implies less mass
dependence to the evolution.  Hence one should regard the detailed
mass dependence of the evolution as a less robust prediction.

The trend to earlier epochs for the molecular gas (middle panel)
is broadly similar to that for the atomic gas, in that it is
increasing at all masses.  There is a steady decrease in $f_{H2}$
with time across all $M_*$ of about $0.2-0.3$ dex between $z=2\rightarrow
0$, with only a very slight trend for more evolution at the highest
masses.  Predicted gas fractions continue to increase at a given
$M_*$ out to $z=4$, so we expect even more gas-rich galaxies at
high masses, but unfortunately even ALMA will have difficulty
measuring the molecular content at these epochs except in the very
largest systems~\citep{Decarli-16}.

Comparing to observations, it appears that \mufasa\ predicts H$_2$
fractions that are too low versus data at $z\sim 1-2$.  CO-based
gas fractions from \citet{Tacconi-13} show a large scatter but
generally lie between $20-40$\% for the most massive galaxies, and
50\% for moderate-mass galaxies.  To lower masses, fractions up to
90\% are inferred for the smallest $z\sim 2$ galaxies by inverting
the \citet{Kennicutt-98} relation~\citep[e.g.][]{Erb-06}.  The dust
continuum-based measures from \citet{Scoville-16} also show typically
molecular fractions of $20-40$\% for main sequence galaxies at
$z\sim 1-2$, and even higher for starbursts.  In contrast, \mufasa\
predicts $z=2$ gas fractions of $\sim 10\%$ for massive ($M_*\sim
10^{11}$) galaxies, and only up to $\sim 40\%$ (relative to the
molecular+stellar mass) for the smallest galaxies that are
well below the what can be probed directly with observations.
Hence in general it appears that \mufasa\ high-$z$ gas fractions
are too low by $\sim\times 2$.  Given the uncertainties this is 
not a gross failure, but it is notable.

Such low high-$z$ molecular gas fractions are predicted in other
simulations and SAMs as well~\citep{Popping-15b,Lagos-15}.  This
may be partially but certainly not completely explained by selection
effects in which targeted CO observations tend to select highly
star-forming (and thus gas-rich) galaxies; \citet{Tacconi-13}
accounted for this and still found generally higher $f_{H2}$ than
predicted here.  Another possibility is that locally-calibrated
CO-to-H$_2$ conversion factors may not be correct at high-$z$; we
will explore this issue in more depth in \S\ref{sec:colf}.  Nonetheless,
at face value it appears that many models including \mufasa\ struggle
to reproduce quite as high gas fractions as inferred for massive
high-$z$ galaxies.

\subsection{\ion{H}{i} mass function}\label{sec:himf}

\begin{figure}
  \centering
  \subfloat{\includegraphics[width=0.45\textwidth]{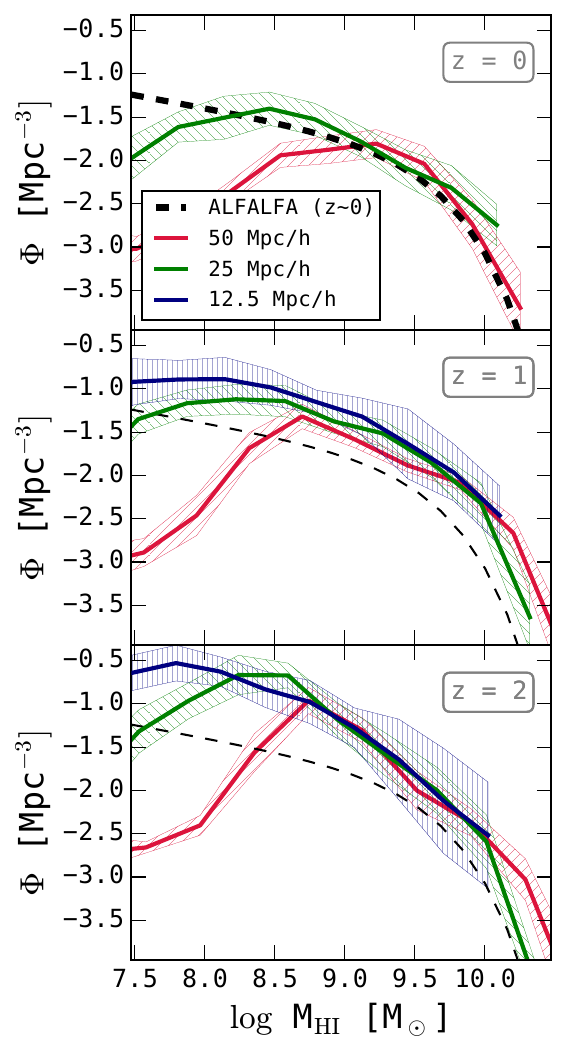}}
  \vskip-0.1in
  \caption{\ion{H}{i} mass functions in the $50\hmpc$ (red solid), $25\hmpc$ (green dashed), 
and $12.5\hmpc$ (blue dotted, for $z\geq1$) \mufasa\ simulations.  Hatched region shows
the cosmic variance computed over 8 sub-octants in each volume.  Black dashed
line shows $z\approx 0$ observations from the ALFALFA survey, reproduced
at $z=1,2$ in order to better depict the evolution.
\mufasa\ does reasonably well reproducing the HIMF over the 
$M_{HI}$ range of convergence; at $z\sim 0$, the $50\hmpc$ box begins to deviate
from the $25\hmpc$ at $M_{HI}\la 10^9M_\odot$, while at higher redshifts this
occurs at $M_{HI}\sim 10^{8.7}M_\odot$.  The HIMF generally increases in amplitude
to higher redshifts, and also steepens noticeably at $z=2$.  There is fair resolution
convergence down to $M_{HI}\sim 10^9M_\odot$ in the $50\hmpc$ volume, and to $\approx 8\times$
lower in the $25\hmpc$ volume.
}
\label{fig:HIMF}
\end{figure}

The \ion{H}{i} mass function (HIMF) combines information from the
galaxy mass function and \ion{H}{i} fractions to provide a complementary
constraint on models.  Observations of the HIMF extend to quite low
masses locally thanks to deep surveys with the Arecibo telescope
such as the \ion{H}{i}-selected The Arecibo Legacy Fast Alfa
survey~\citep[ALFALFA;][]{Haynes-11} and the stellar mass-selected
Galex Arecibo SDSS Survey~\citep[GASS;][]{Catinella-10}.  However,
the sensitivity of current instrumentation precludes characterisation
of the HIMF at significantly higher redshifts.  The SKA and its
precursors aim to improve on this, and hence predictions for the
evolution of the HIMF are useful for quantifying expectations for
upcoming surveys such as the Looking At the Distant Universe with
the MeerKAT Array~\citep[LADUMA;][]{Holwerda-12} survey.

Figure~\ref{fig:HIMF} shows the predicted HIMF from \mufasa, showing
the three volumes in different colours with cosmic variance (shading)
estimated as before, from the variance of the HIMF in each of the
8 sub-octants within each simulation volume.  As discussed in
Paper~I, this is likely to somewhat underestimate the true cosmic
variance.  The three panels show the HIMF at $z=0,1,2$ from top to
bottom.  The ALFALFA mass function at $z\approx 0$ is shown as the
thick black dashed line, and this is reproduced in the other redshift
panels to better visualise the amount of evolution in the models;
however, direct comparison to \mufasa\ should only be made at $z=0$.

The top panel shows that \mufasa\ provides a reasonable match to
the observed HIMF, in the resolved range.  At $M_{HI}\la 10^9
M_\odot$, the $50\hmpc$ volume shows a departure from the data, but
the higher resolution run continues unabated, suggesting that the
turnover at low masses is an artifact of numerical resolution.
Indeed, if one combines the stellar mass resolution limit of
$10^{8.7}M_\odot$ with the fact that galaxies at that $M_*$ have
an \ion{H}{i} fraction of around two (Figure~\ref{fig:fgas}), this
suggests that galaxies with $M_{HI}\la 10^9M_\odot$ will suffer
from incompleteness in our simulations.  The $25\hmpc$ volume extends
another factor of almost 8 lower in mass before turning over, as
expected from its $8\times$ higher mass resolution.

The agreement of \mufasa\ with both the stellar mass function
(Paper~I) and the HIMF is an important success.  Previous simulations
by \citet{Dave-13} also showed good agreement with data for both,
even when subdivided into stellar mass bins \citep{Rafieferantsoa-15}.
The EAGLE simulation likewise shows good agreement with
both~\citep{Crain-16}, including subdivided into $M_*$
bins~\citep{Bahe-16}.  However, semi-analytic models constrained
to match the stellar mass function don't necessarily agree well
with the HIMF~\citep[e.g.][]{Benson-14}.  The SAMs of \citet{Popping-14}
do fairly well at $M_{HI}\ga 10^9M_\odot$, but predict a significant
upturn to lower masses that is not observed.  Such an upturn is
also seen in the older Overwhelmingly Large Simulations (OWLS) HIMF
as well~\citep{Duffy-08}.  The semi-empirical model of \citet{Popping-15}
likewise produces a steep faint end of the HIMF, deviating strongly
at $M_{HI}\la 10^9M_\odot$.  In simulations such as \mufasa,
\ion{H}{i} represents a transient reservoir of cold gas infalling
into a galaxy, as demonstrated in \citet{Crain-16}; such a dynamic
origin suggests that fully dynamical models are best suited to make
predictions for the nature and evolution of \ion{H}{i} in galaxies.
It appears that the low-mass ($M_{HI}\la 10^9M_\odot$) HIMF may be
a key discriminant for the dynamics of gas infall.

Looking at the evolution to $z=1$, we see that the HIMF is best
described by an overall increase in the mass of \ion{H}{i} in each
galaxy by a factor of $\sim\times 2-3$, particularly at the massive
end.  This is consistent with the evolution seen in
Figure~\ref{fig:fgasevol}.  This is good news for surveys such as
LADUMA that will probe the bright end of the HIMF at these redshifts;
in future work we will make more specific predictions for LADUMA.
Interestingly, this is somewhat contrary to the trend predicted by
our previous simulations in \citet{Dave-13}, which showed a steepening
of the HIMF to higher redshifts, but the massive end was generally
unchanged or lowered.  This is because the \ion{H}{i} fraction in
the \citet{Dave-13} simulations was invariant with redshift, whereas
in \mufasa\ galaxies are substantially more \ion{H}{i}-rich,
particularly at high masses.

From $z=1$ to 2, the main trend is that the HIMF is steeper at low
masses, while the massive end does not evolve significantly.  This
is driven by the steepening of the stellar mass function, since the
\ion{H}{i} fraction if anything has a shallower trend with stellar
mass at higher redshifts (Figure~\ref{fig:fgasevol}).  This general
trend agrees better with that in \citet{Dave-13}.

Overall, the HIMF in \mufasa\ at $z=0$ is a reasonable match to
observations, even though the dynamic range is limited compared to
other simulations such as EAGLE.  By combining various box sizes,
we can span a similar dynamic range, and the HIMF shows good
resolution convergence in the overlapping \ion{H}{i} mass range.
\mufasa\ predicts a noticeably higher HIMF at $z\sim 1$, and then
a steepening trend to $z\sim 2$, which must await future SKA and
pathfinder telescope data for testing.

\subsection{CO luminosity function}\label{sec:colf}

\begin{figure}
  \centering
  \subfloat{\includegraphics[width=0.45\textwidth]{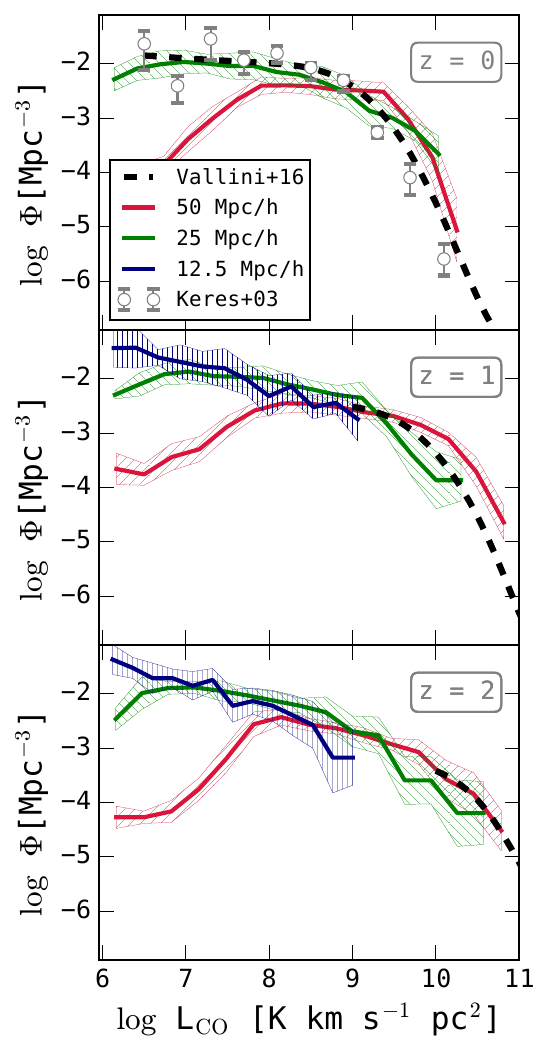}}
  \vskip-0.1in
  \caption{CO luminosity functions in the $50\hmpc$ (red solid), $25\hmpc$ (green dashed),
and $12.5\hmpc$ (blue dotted, for $z\geq1$) \mufasa\ simulations.  Hatched region shows
the cosmic variance computed over 8 sub-octants in each volume.  
We compute the CO1-0 luminosity from our predicted H$_2$ mass based on 
the prescription in \citet{Narayanan-12} derived from
zoom simulations coupled with CO radiative transfer.  
We show observations
from \citet{Vallini-16} at $z=0, 1, 2$ as the black dashed lines, down to their
approximate completeness limit at each redshift; note that these are based on
an $L_{IR}$-to-$L_{CO}$ conversions.  Resolution convergence amongst the
volumes is generally quite good.
}
\label{fig:COLF}
\end{figure}

The mass function of molecular gas is more complicated to determine
than that of atomic gas, since observations typically do not
directly trace H$_2$ but rather some proxy such as CO.  For ordinary
(non-starburst) galaxies, canonically the best proxy for H$_2$ is the
$J=1-0$ rotational transition of CO.  Nonetheless, this still
requires a conversion factor ($X_{CO}$) to obtain the H$_2$ mass,
and the dependence of $X_{CO}$ on the intrinsic properties of
galaxies such as star formation rate and metallicity is uncertain.
This becomes particularly problematic at high redshifts, where the
ISM conditions in typical main sequence galaxies vary substantially
from that today.

A typical assumption is that galaxies that are near the main sequence
have ``Milky Way-like" $X_{CO}\approx 4$, whereas starbursts have
$X_{CO}\approx 0.8$~\citep{Tacconi-13}.  However, substantial work
has gone into predicting $X_{CO}$ based on galaxy properties from
detailed simulations, yielding a continuous rather than bimodal trend.
In particular, \citet{Narayanan-12} used zoom simulations together with a
CO line radiative transfer code to develop an approximate fitting function for
$X_{CO}$ as a function of H$_2$ surface density and metallicity:
\begin{equation}
X_{CO} = {1.3\times 10^{21}\over Z' \times \Sigma_{H2}^{0.5}}
\end{equation}
where $Z'$ is the metallicity in solar units.  

Here, we use this formula to compute $X_{CO}$ individually for each
galaxy, obtaining $\Sigma_{H2}$ by dividing the H$_2$ half-mass of each galaxy 
by the area computed from the H$_2$ half-mass radius.
Using this $X_{CO}$, we then convert our
simulated H$_2$ masses into CO luminosities ($L_{CO}$), which can
be compared more directly against observations.  In this way, we
specifically account for the metal and gas content evolution in
CO-to-H$_2$ conversions when comparing to observations.  This is
analogous to the approach in \citet{Narayanan-12b}, except that
here we convert simulated galaxies to get $L_{CO}$, while they took
the converse approach of converting observations into $M_{H2}$ to
compare with models.  However, we will see that our conclusions are
similar.

Figure~\ref{fig:COLF} shows the CO luminosity function (COLF) from our
\mufasa\ simulations, showing once again our available simulation
volumes at each redshift $z=0,1,2$ (top to bottom).  At $z=0$, it
is possible to directly observe CO~1-0 down to very low $L_{CO}$,
and such observations by \citet{Keres-03} are shown as the data
points.  To higher redshifts, blind CO surveys where the survey
volume can be robustly estimated are difficult, so one typically
uses another proxy for this.  The dashed lines show observations
from \citet{Vallini-16}, which used far-infrared luminosity as a
proxy for CO luminosity; at $z=0$, they agree with the \citet{Keres-03}
data.  At higher redshifts, we plot their observations down to their
approximate completeness limit.  We note that recent direct CO
measures from ALMA by \citet{Decarli-16} indicate a somewhat higher
number of high-$L_{CO}$ objects at $z\sim 2$ than \citet{Vallini-16},
but the statistics are small and the cosmic variance is large, so
the discrepancy is only marginally significant.

At $z=0$, \mufasa\ generally predicts a reasonable COLF, with a
hint of an excess at high $L_{CO}$.  The $50\hmpc$ volume shows a
turnover at low-$L_{CO}$ owing to numerical resolution, while the
$25\hmpc$ continues to agree well with the observations down to the
lowest probed $L_{CO}$.  The observations of \citet{Vallini-16}
generally find an increase in the number of high-$L_{CO}$ galaxies
with redshift, and the simulations follow this trend, generally
agreeing with data with still a hint of a high-$L_{CO}$ excess.  By
$z=2$ the observations only probe the brightest CO galaxies, where
only the $50\hmpc$ volume has comparably bright systems, but these
are in very good agreement with the data.

It is interesting that despite the relatively mild evolution of
H$_2$ fractions in Figure~\ref{fig:fgasevol} and a putative
undeprediction of $f_{H2}$ at $z\sim 2$, \mufasa\ reproduces well
the evolution of the COLF out to $z=2$, and shows significantly
more high-$L_{CO}$ objects at high redshifts.  This suggests that
using a physically-motivated prescription for converting CO into
H$_2$ (or vice versa) can lead to inferring a different amount of
evolution in the gas fractions, and in general could potentially reconcile the
relatively low amount of evolution in simulations versus the stronger
evolution inferred using standard assumptions regarding $X_{CO}$;
this broadly echoes the conclusions of \citet{Narayanan-12b}.
Generically, metallicity-dependent $X_{CO}$ prescriptions such as
\citet{Narayanan-12} and \citet{Feldmann-12} tend to predict more
H$_2$ at high masses and less at low masses owing to enhanced H$_2$
production at high metallicities, which serves to increase the
bright end of the COLF and flatten the faint end~\citep{Popping-14}
thus yielding better agreement with the COLF.  Empirical
luminosity-dependent $X_{CO}$ calibrations have a qualitatively
similar effect~\citep{Boselli-14}.  Hence \mufasa\ may be plausibly
reproducing the evolution of the molecular gas content in galaxies
in spite of its modest evolution of $\sim\times 2$ in the H$_2$
content at a fixed $M_*$ out to $z\sim 2$.

\begin{figure}
  \centering
  \subfloat{\includegraphics[width=0.45\textwidth]{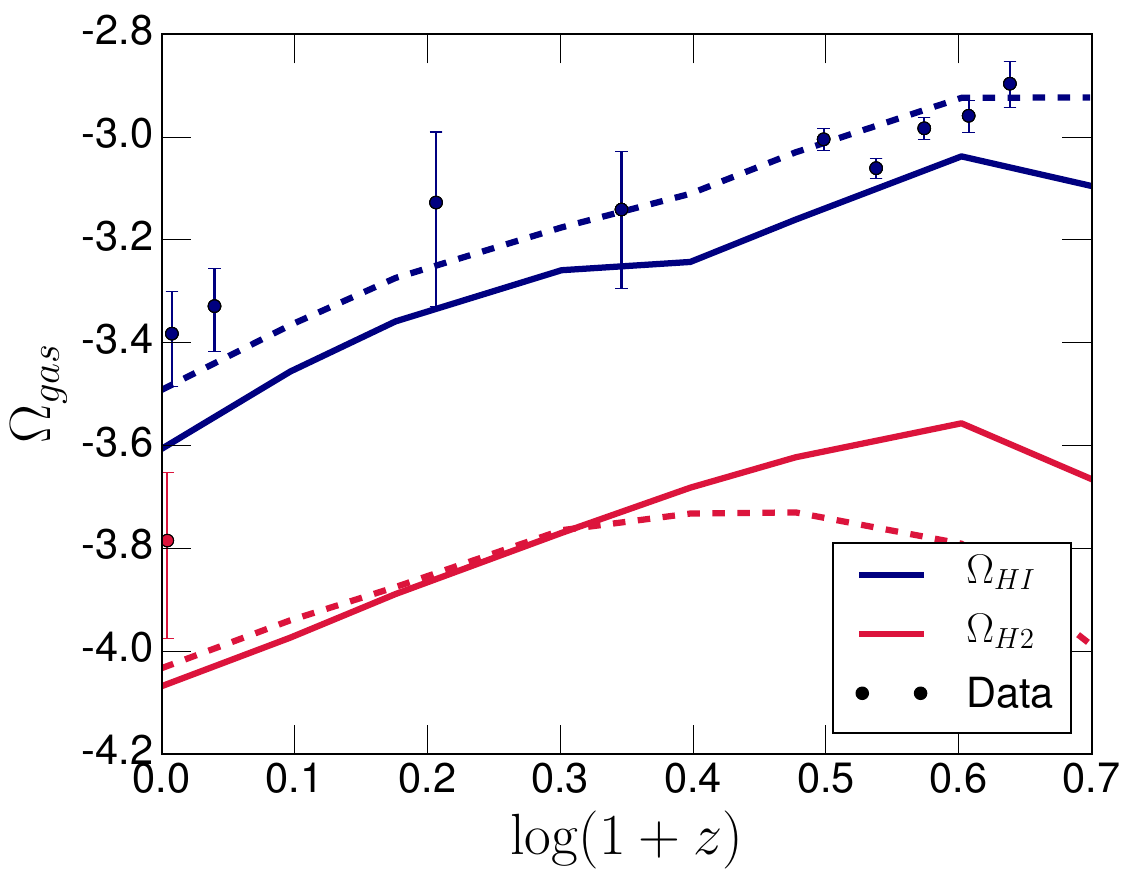}}
  \vskip-0.1in
  \caption{Cosmic mass density in \ion{H}{i} (blue) and H$_2$ (red) as a function
of redshift, in our $50\hmpc$ (solid) and $25\hmpc$ (dashed) \mufasa\ simulations.
$\Omega_{HI}$ observations are shown from \citet[$z<0.1$]{Delhaize-13},
\citet[$0.5<z<1.3$]{Rao-06}, and \citet[$2.1<z<3.35$]{Noterdaeme-12}.  The
predicted trend of approximately $\Omega_{HI}\propto (1+z)^{0.74}$ is a good match to
the compiled observations, as is the normalization although it is
somewhat sensitive to numerical resolution.  $\Omega_{H2}$ shows a 
similar redshift trend as $\Omega_{HI}$, which is substantially slower
than the evolution of the cosmic SFR density as shown in Figure~3 of Paper~I.
}
\label{fig:omegagas}
\end{figure}

\subsection{Cosmic gas mass evolution}\label{sec:omegagas}

A synthesis of all the above evolutionary measures is provided in
the evolution of the global cosmic gas density, typically parameterised
in units of the critical density (i.e. as $\Omega_{\rm gas}$).  The
slow evolution of $\Omega_{HI}$ relative to the overall cosmic star
formation rate density has been noted as evidence that \ion{H}{i}
is not directly physically associated with star formation, while
the more rapid evolution of H$_2$ fractions can explain at least
part of the rapid evolution in the cosmic SFRD.  However, such
interpretations are complicated by detailed assumptions regarding
$X_{CO}$ as discussed in the previous section, and how \ion{H}{i}
gas traces galaxies.  Here we examine predictions for the evolution
of the cosmic \ion{H}{i} and H$_2$ mass densities, in the context
of the evolutionary trends we have discussed above.

Figure~\ref{fig:omegagas} shows the evolution of the cosmic density
in atomic gas (blue) and molecular gas (red) as a function of
$\log(1+z)$.  Solid and dashed lines show the results from our
$50\hmpc$ and $25\hmpc$, respectively.  This is obtained by summing
over all SKID-identified galaxies; using instead the sum over all
\ion{H}{i} or H$_2$ in the volume (which includes the IGM) makes a
negligible difference.

Data points with the blue error bars correspond to various observational
measures of $\Omega_{HI}$: From 21cm emission \citep[$z<0.1$]{Delhaize-13},
using \ion{Mg}{ii} absorbers as a proxy for
DLAs~\citep[$0.5<z<1.3$]{Rao-06}, and DLA absorbers selected from
the Sloan Digital Sky Survey~\citep[$2.1<z<3.35$]{Noterdaeme-12}.
For H$_2$, no data is shown; at $z=0$, \citet{Keres-03} inferred
$\Omega_{H2}\approx 2\times 10^{-4}$ which is above the predictions,
but given the good agreement shown versus the $z\approx 0$ COLF
from the previous section, this could be subject to uncertainties
regarding $X_{CO}$.

$\Omega_{HI}$ roughly follows a power law in $(1+z)$; a best-fit
relation to the $50\hmpc$ run is given by $\Omega_{HI} =
10^{-3.53}(1+z)^{0.74}$, and is higher in amplitude by 20\% for the
$25\hmpc$ volume.  Generally, this provides a good fit to the trend
seen in the compilation of observations from various sources and
techniques, particularly for the higher-resolution volume.  The
difference between the volumes, while only about 0.1~dex, nonetheless
suggests that there is suboptimal resolution convergence in this
quantity, likely driven by the fact that the $50\hmpc$ volume does
not resolve many low-$M_{HI}$ galaxies as seein in Figure~\ref{fig:HIMF}.
Semi-analytic models tend to predict that $\Omega_{HI}$ rises
somewhat out to intermediate redshifts, but then falls at $z\ga
1-2$~\citep{Obreschkow-09,Popping-15}, in clear disagreement with
a continued rise in $\Omega_{HI}$ out to $z\sim 3.5$.  Hence the
broad agreement in the redshift evolution of $\Omega_{HI}$ is highly
encouraging, and suggests that \ion{H}{i} in and around galaxies
is being viably modeled by \mufasa\ across a range of epochs.

In contrast, the evolution of $\Omega_{H2}$ is predicted to be
substantially slower than often believed.  \mufasa\ predicts
essentially the same redshift evolution for $\Omega_{H2}$ as for
$\Omega_{HI}$, with an increase of a factor of $\times 3$ from
$z=0\rightarrow 3$.  The SAMs of \citet{Lagos-11}
predict almost no evolution for $\Omega_{HI}$, but a $\times 7$
increase from $z=0\rightarrow 3$ for $\Omega_{H2}$.  More recently,
\citet{Lagos-15} found slower evolution of $\Omega_{H2}$ in the
EAGLE simulation, more similar to \mufasa.  Observations cannot yet
clearly distinguish between these predictions.

In summary, \mufasa\ predicts mild evolution in both the total
\ion{H}{i} and H$_2$ cosmic mass densities, scaling approximately
as $(1+z)^{0.8}$.  Such a scaling roughly follows from the simple
equilibrium model arguments outlined at the start of this section.
The evolution of $\Omega_{HI}$ is in good agreement with observations,
but the predictions for $\Omega_{H2}$ are not currently robustly
testable.  As CO and far-infrared surveys improve with ALMA and
other facilities, such constraints will provide important tests of
these and other models.

\section{Fluctuations around scaling relations}\label{sec:dev}

In the prevalent baryon cycling paradigm, quasi-continuous gas
inflows drive galaxy growth, modulated by feedback~\citep{Somerville-15}.
The net result is that galaxies live on fairly tight scaling relations
between stellar mass, star formation rate, metallicity, and gas
content~\citep[e.g.][]{Finlator-08,Dave-12,Lilly-13,Lagos-15}.  Fluctuations
in the inflow rate owing to e.g. mergers can cause fluctuations
around these scaling relations.  Indeed, inflow fluctuations owing
to stochastic dark matter infall alone yield a scatter that is in
good agreement with the observed scatter in the SFR$-M_*$
relation~\citep{Forbes-14,Mitra-16}.

In addition to scatter in SFR, such fluctuations also give rise to
correlated scatter in the metallicity and gas content.  For instance,
a boost in inflow will enhance the gas content, lower the metallicity,
while boosting the SFR owing to the abundance of fresh fuel.  Hence
one expects that, at a given $M_*$, high gas content should correlate
with low metallicity and high SFR.  This results in second-parameter
dependences with SFR in the scatter around these scaling relations.
In this section we quantify these second-parameter dependences in
\mufasa, which provides predictions for baryon cycling that are
testable with current and future observations.

Figures~\ref{fig:mzr} and \ref{fig:fgas} already showed clear
second-parameter dependences on the SFR in \mufasa\ galaxies:
Galaxies that have higher SFR for their $M_*$ also have lower
metallicities and higher gas fractions in both \ion{H}{i} and H$_2$.
To quantify this, we use ``deviation plots," i.e. we plot the
deviation away from the mean scaling relation in two quantities
versus $M_*$ against each other.  This isolates the second-order
aspects of baryon cycling-driven galaxy evolution by directly
quantifying how fluctuations drive correlated scatter, while removing
the dependence on the overall inflow rate that sets the first-order
(mean) scaling relationship between quantities.

As an example, in order to make a deviation plot for sSFR vs. \ion{H}{i},
we begin with the sSFR$-M_*$ and $f_{HI}-M_*$ relations.  For each
galaxy, we then compute the difference between $\log$~sSFR of that
galaxy and the median of all galaxies' $\log$~sSFR at that $M_*$; we call
this $\Delta\log$~sSFR.  Similarly, we compute the difference between
$\log f_{HI}$ for that galaxy and the median $\log f_{HI}$ at that galaxy's
$M_*$; this is $\Delta \log f_{HI}$.  We can analogously compute $\Delta
\log f_{H2}$ and $\Delta\log Z$ for the molecular gas and 
metallicity, respectively.  Note that here we are always using $M_*$
as our independent variable, because this quantity is stable on the
(relatively) short timescales over which deviations are occuring;
in principle, it is possible to use any property as the independent
variable, but we leave such explorations for future work.

\begin{figure*}
  \centering
  \subfloat{\includegraphics[width=0.85\textwidth]{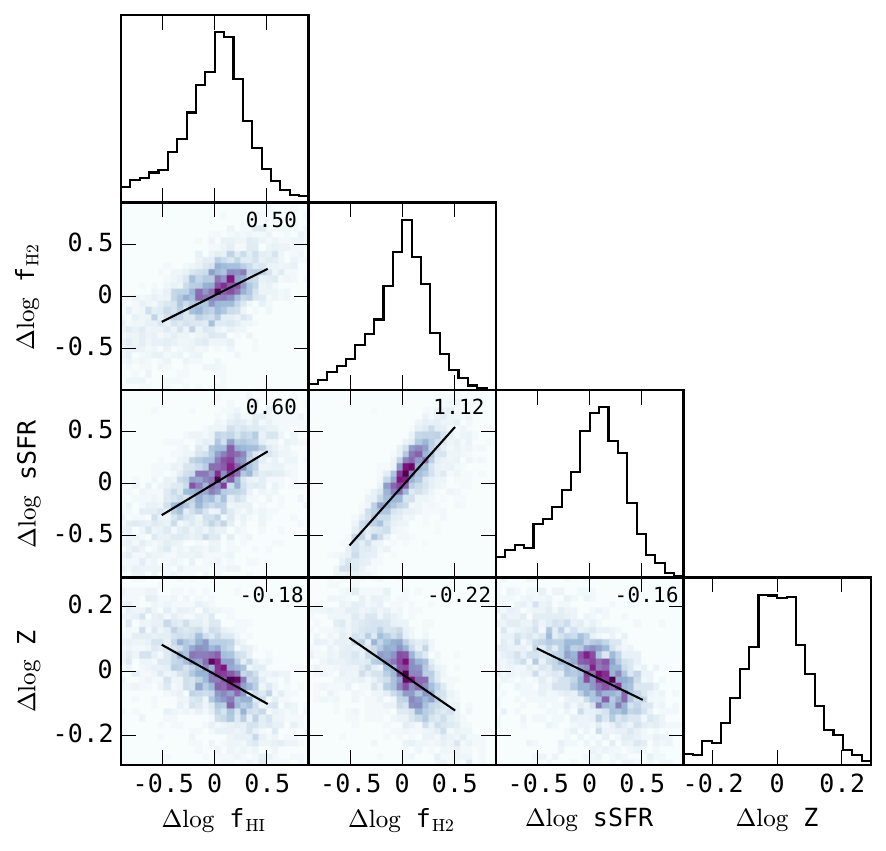}}
  \vskip-0.1in
  \caption{Plots showing the deviation from mean scaling relations versus $M_*$ in
our $50\hmpc$ volume at $z=0$ for four
quantities (in logarithm): \ion{H}{i} fraction, H$_2$ fraction, specific SFR, and metallicity.  
Scatter plots show these deviations plotted against each other, depicting how
fluctuations in these quantities are correlated.  The trends can be reasonably
represented by the best-fit power laws shown as the solid lines, with the slope
indicated in the upper right of each panel.  Overall, galaxies at a given $M_*$
with higher sSFR have higher \ion{H}{i} and H$_2$ fractions and lower metallicity.  The panels
along the diagonal show histograms of the scatter around each scaling relation.
While generally Gaussian, there is a tail to low sSFR and gas content arising from 
green valley galaxies.
}
\label{fig:dev}
\end{figure*}

Figure~\ref{fig:dev} shows 2-D histograms of the deviations $\Delta\log
f_{HI}$, $\Delta\log f_{H2}$, $\Delta\log$sSFR, and $\Delta\log Z$
plotted against each other.  Only star-forming galaxies are included,
and for simplicity we only show the $50\hmpc$ volume at $z=0$ but
the trends are similar in the other volumes.  The panels along the
diagonal show the histograms of deviation values for each quantity,
which illustrate the shape of the scatter around the median scaling
relation versus $M_*$.  The solid line in each panel shows the
best-fit power law to the deviations shown, and the number in the
upper right corner is the best-fit slope.

Figure~\ref{fig:dev} at its most basic level shows that deviations
in the SFR, \ion{H}{i}, and H$_2$ all correlate positively with
each other, while metallicity deviations (bottom row) anti-correlate
with all of the others.  This quantifies the amount by which galaxies
that lie above the mean MZR also tend to lie below the mean relations
in sSFR, $f_{HI}$, and $f_{H2}$ vs.  $M_*$.  Such trends arise
naturally in a ``gas regulator" type model, which is an ISM
mass-balance formalism in which the gas content is allowed to
vary~\citep{Lilly-13}.

The slope of the best-fit line contains information about how well
quantities track each other.  For instance, consider $\Delta\log
f_{H2}$ vs. $\Delta\log$~sSFR: The slope is close to linear, which
means that fluctuations in H$_2$ are directly tracking fluctuations
in SFR.  This is unsurprising, since in our simulations it is assumed
that the star formation rate of any given gas element is proportional
to its $f_{H2}$; nonetheless, it is not trivial that this translates
into a similar trend in galaxy-integrated quantities.  The slope
versus $\Delta\log f_{HI}$, in contrast, is somewhat sublinear for
$\Delta\log f_{H2}$ and $\Delta\log$sSFR, indicating that fluctuations
in H$_2$ and sSFR do not perfectly reflect fluctuations in \ion{H}{i}.

The deviations in metallicity versus the MZR, i.e. the plots along
the lowermost row, have garnered much attention in the literature.
For instance, the panel showing $\Delta\log$~sSFR vs. $\Delta\log
Z$ (lower rightmost) corresponds to the FMR, showing that galaxies
with higher SFR at a given $M_*$ have lower $Z$.  The best-fit line
has a slope of $-0.16$, which represents a higher-order testable
prediction of \mufasa's ability to depict the fluctuations in baryon
cycle that give rise to scatter around the scaling relations.

The bottom leftmost panel corresponds to the observational trend
noted by \citet{Hughes-13}, \citet{Lara-Lopez-13}, and \citet{Bothwell-13},
which the latter dubbed the \ion{H}{i}-FMR: Galaxies with higher
\ion{H}{i} content at a given $M_*$ are seen to have lower
metallicities.  Resolved spectroscopy by \citet{Moran-12} indicated
that the excess in \ion{H}{i} tends to be accompanied by a drop in
the outer metallicity, strongly suggesting that this trend is driven
by accretion in the outskirts of galaxies.  \mufasa\ predicts a
slope for $\Delta\log f_{HI}$ vs.  $\Delta\log Z$ of $-0.18$, similar
to but slightly stronger than that vs. $\Delta\log$sSFR.
\citet{Robertson-13} measured this deviation slope to be $-0.41\pm
0.14$ for field galaxies ($-0.31$ for cluster galaxies).  This is
steeper than our current predictions, but this was done at a fixed
sSFR rather than $M_*$, which likely accounts for some of the
difference.  Metallicity is formally most strongly tied to H$_2$;
the slope of this deviation relation is $-0.22$.

One can also examine the spread of points around the best-fit linear
relation within the deviation plots.  This is another measure of
how tightly any given two quantities flucutate.  One can quantify
this by measuring the mean deviation in, say, metallicity, from the
best-fit relations involving $\Delta\log Z$.  For the metallicity
relations, the mean departure in $\Delta\log Z$ is $0.062, 0.061,
0.064$ with respect to $\Delta\log f_{HI}$, $\Delta\log f_{H2}$,
and $\Delta\log$sSFR, respectively.  This again suggests, at a very
marginal level, that metallicity more strongly follows \ion{H}{i}
than sSFR, which is a conclusion also reached in observational
analysis by \citet{Bothwell-13}.  Still, metallicity tracks H$_2$
slightly better than either of these quantities~\citep[as also found
by][]{Lagos-15}.

Finally, the diagonal panels show the scatter of each quantity
around the mean scaling relation versus $M_*$.  The shape is generally
Gaussian, with a spread that is slightly smaller in H$_2$ relative
to sSFR and \ion{H}{i}.  Metallicity has quite small scatter,
consistent with $\sim 0.1$~dex as observed~\citet{Tremonti-04}.  In
detail there is a longer tail to low-$\Delta$sSFR and correspondingly
low gas fraction deviations, which arises from galaxies on their
way to quenching.

This deviation plot represents the global view over all galaxies
down to the resolution limit of our $50\hmpc$ volume at $z=0$.
Clearly, it is instructive to examine this plot using galaxies
binned by mass, or colour, or at different redshifts.  We do not
show this here, but we have checked that the trends depicted in
Figure~\ref{fig:dev} are generally well-converged with resolution
in the overlapping mass range, and they are qualitatively similar
at higher redshifts.  One can also examine trends by fixing other
quantities besides $M_*$, such as SFR.  In future work, we will
explore the implications of these deviation plots in terms of baryon
cycling, and present more detailed comparisons to relevant observations.

In summary, deviations plots quantify how galaxies respond to
fluctuations in the baryon cycle.  By examining only the departures
around the mean relations (with respect to $M_*$), we remove the
``first-order" component of galaxy growth (along with many associated
systematics) and isolate the impact
of ``second-order" fluctuations on observable quantities.  We thus
quantify the correlation in scatter among these various quantities,
thereby presenting a new and higher-order test of galaxy formation
models.  While there currently exist various forms of observational
characterisations for these trends, we plan to conduct a more
thorough and direct comparison to data regarding second-parameter
trends for both gas fractions and metallicities in future work.

%

\section{Summary}\label{sec:summary}

We have presented predictions of the \mufasa\ simulations and
compared to observations of the star formation rate, metal, and gas
content of galaxies.  \mufasa\ uses state of the art feedback modules
and hydrodynamics methodology taken from high-resolution zoom
simulations and analytic models.  To further extend our dynamic
range we employ several simulations using identical input physics
but varying in volume (box sizes of $50, 25, 12.5\hmpc$), and check
that generally simulations at different numerical resolution make
similar predictions in their overlapping mass ranges.

Following on Paper~I where we showed that \mufasa\ performed
creditably at reproducing the observed stellar mass function over
a range of cosmic epochs, here we further show that it also fares
well against a number of other key barometers, including several
that have not been examined extensively versus previous models such
as the specific star formation rate function.  We also make novel
and testable predictions for the correlations in the fluctuations
around mean scaling relations in SFR, metallicity, and gas content,
as a direct means to quantify how galaxies respond to fluctuations
in the baryon cycle.

Our main results are summarised as follows:
\begin{itemize}

\item The star formation rate function in \mufasa\ shows a Schechter
shape with a relatively shallow faint end, in broad agreement with
observations out to $z\sim 2$, albeit with a hint that \mufasa\
overpredicts high-SFR galaxies.  This is curious given that in
Paper~I we demonstrated that \mufasa\ matches the stellar mass
function well but strongly underpredicts the $z\sim 2$ sSFR$-M_*$
relation, which implies that, if anything, \mufasa\ should {\it
under}predict the SFR function.  This highlights the continued
difficulty in reconciling current SFR measures during Cosmic Noon
with models and, in some cases, among the various data sets themselves.

\item The specific SFR function provides a more detailed test of
how well models reproduce the scatter around the main sequence.
\mufasa\ reproduces the observed sSFR function at low-$z$ quite
well, indicating that this simulation is nicely reproducing the
number of galaxies in the green valley, is correctly capturing the
spread around the main sequence, and is not missing a large population
of starbursts.  At $z\sim 1$, the entire sSFR function is shifted
by $\sim\times 2$ with respect to observations although the shape
matches well, reiterating the result from Paper~I showing that the
mean sSFR at that epoch is underpredicted by a similar factor.

\item The mass-metallicity relation shows a reasonable low-mass
slope and amplitude versus observations at both $z\approx 0$ and
2.  
In contrast, $M_*\ga 2\times 10^{10}M_\odot$ star-forming
galaxies at $z=0$ continue to show a strong rise in the MZR that
does not agree well with observations, and then abruptly flattens
at roughly the appropriate metallicity.  We conjecture that wind
recycling, which plays a key role in setting the MZR at these masses,
may be too vigorous in our simulations at these masses, or else
these galaxies should have a metal loading factor above our assumed
value of unity.  Finally, the MZR clearly shows a second-parameter
trend such that galaxies with high SFR at a given $M_*$ have lower
metallicity.

\item \mufasa\ directly tracks molecular gas, hence we can separate
the gas content into atomic, molecular, and ionised.  \mufasa\ well
reproduces observations of the total cold gas fraction (\ion{H}{i}+H$_2$)
as a function of $M_*$, and provides a fair match to the \ion{H}{i}
fraction individually, with a notable deficit at high masses in
only our lowest-resolution run.  Like with the metallicity, \ion{H}{i}
and H$_2$ content also show a second parameter trend that galaxies
with high SFR tend to have higher gas fractions.

\item Gas fractions are broadly predicted to increase with redshift,
which at least qualitatively agrees with observations.  However,
the predicted rate of evolution for H$_2$ ($\sim\times 2-3$ out to $z\sim 2$)
is slower than canonically observed for molecular gas.  H$_2$ evolves
similarly across all masses, while \ion{H}{i} evolves slightly
faster at higher masses; there is an order of magnitude more
\ion{H}{i} in a $M_*=10^{11}M_\odot$ galaxy at $z\sim 1-2$ versus
today.

\item As a result of the rapid \ion{H}{i} evolution at high masses,
the bright end of the \ion{H}{i} mass function evolves fairly rapidly
as well.  The predicted HIMF agrees well with observations at $z\sim
0$, and evolves upwards at all masses by $\sim\times 2-3$ by $z=1$.
At $z=2$ we predict a steeper faint end, although this may not be
accessibly observationally in 21~cm prior to the full SKA.

\item In order to explore the potential discrepancy in molecular
gas evolution further, we use the simulation-based prescription
from \citet{Narayanan-12} to convert \mufasa\ molecular gas masses
to a CO luminosity based on the metallicity and molecular gas
content, and compare to inferred CO luminosity functions observed
out to $z\sim 2$.  We find surprisingly good agreement at all masses
for the COLF, despite the mild evolution in $f_{H2}$.  This highlights
that systematic uncertainties in $X_{CO}$ can be a overriding factor
in making robust comparisons to molecular gas content data at
intermediate redshifts.

\item The cosmic mass density in \ion{H}{i} is predicted to evolve
mildly upwards out to high-$z$, such that $\Omega_{HI}\propto
(1+z)^{0.7-0.8}$.  This evolution is in good agreement with
observations from a variety of techniques.  The amplitude is somewhat
sensitive to resolution, and our $25\hmpc$ volume has 10--20\% higher
$\Omega_{HI}$ than our $50\hmpc$ cube, which agrees slightly better
with data.

\item In accord with our predicted mild evolution for $f_{H2}$, we
also predict mild evolution for $\Omega_{H2}$, with a similar
redshift scaling as $\Omega_{HI}$ but lower by $\sim\times 3$.  We
note that this is much less steep than the evolution of the cosmic
SFR density; it has been suggested that the drop in cosmic SFR owes
directly to the drop in molecular gas mass, but \mufasa\ does not
support this intepretation, as the cosmic SFRD (see Paper~I) evolves
significantly more rapidly than $\Omega_{H2}$.

\item An independent and higher-order test of models is whether
they reproduce the observed scatter around the mean scaling relations.
We make predictions for this using deviation plots, where we correlate
the deviations for each galaxy relative to scaling relations in
sSFR, metallicity, $f_{H2}$, and $f_{HI}$ versus $M_*$.  We show
that \mufasa\ qualitatively reproduces observed trends that indicate
that at a given $M_*$, galaxies with high SFR have low metallicity
and high gas content.  We make predictions for the power-law slopes
between deviations in sSFR, metallicity, $f_{H2}$, and $f_{HI}$
that can be tested against observations.

\end{itemize}

As in Paper~I, \mufasa\ continues to demonstrate good agreement
with now a wider range of galaxy observables across cosmic time,
indicating that it provides a viable platform to study the physics
of galaxy evolution in a cosmological context on $\ga$kpc scales.
This implies, among other things, that employing scalings taken
from the FIRE simulations into cosmological-scale runs satisfyingly
reproduces some of the same data-concordant trends as individual
FIRE zoom runs.  Our current heuristic quenching model seems to
populate the green valley and lower their gas contents relative to
blue cloud galaxies approximately as observed, though we are working
towards a more self-consistent black hole growth and feedback model
that may substantially impact these predictions particularly at the
massive end.  \mufasa's successes showcase the emerging promise of
cosmologically-situated galaxy formation simulations in helping to
understand the Universe as mapped through large-scale multi-wavelength
galaxy surveys probing the various constituents of galaxies back
to early epochs.

 \section*{Acknowledgements}
The authors thank D. Angl\'es-Alc\'azar, F. Durier, K.  Finlator,
S. Huang, N. Katz, D. Narayanan, and R. Somerville for helpful conversations and
comments.  RD, MR, and RJT acknowledge support from the South African
Research Chairs Initiative and the South African National Research
Foundation.  Support for MR was also provided by the Square Kilometre
Array post-graduate bursary program.  Support for RD was provided
by NASA ATP grant NNX12AH86G to the University of Arizona.  Support
for PFH was provided by an Alfred P. Sloan Research Fellowship,
NASA ATP Grant NNX14AH35G, and NSF Collaborative Research Grant
\#1411920 and CAREER grant \#1455342.  The \mufasa\ simulations
were run on the Pumbaa astrophysics computing cluster hosted at the
University of the Western Cape, which was generously funded by UWC's
Office of the Deputy Vice Chancellor.  These simulations were run
with revision e77f814 of {\sc Gizmo} hosted at {\tt
https://bitbucket.org/rthompson/gizmo}.

\end{document}